\documentclass[iop]{emulateapj}
\pdfoutput=1
\usepackage{apjfonts}
\usepackage{subfigure}
\usepackage{graphicx}
\usepackage{color}

\newcommand{\ergssec}{~ergs~s$^{-1}$}
\newcommand{\ergss}{~ergs~s$^{-1}$}
\newcommand{\kms}{~km~s$^{-1}$}
\newcommand{\msun}{{M$_{\odot}$}}
\newcommand{\dg}{$^{\circ}$}

\bibpunct{(}{)}{,}{a}{}{}

\def\gtrsim{\mathrel{\hbox{\rlap{\hbox{\lower4pt\hbox{$\sim$}}}\hbox{$>$}}}}

\shorttitle{Characterizing ``Radio Mode'' AGN Outbursts}
\shortauthors{Forman et al.}

\begin{document}

\title{PARTITIONING THE OUTBURST ENERGY  OF A LOW EDDINGTON ACCRETION
  RATE AGN AT THE CENTER OF AN ELLIPTICAL GALAXY: THE RECENT 12 MYR
  HISTORY OF THE SUPERMASSIVE BLACK HOLE IN M87}

\author{W.~Forman\altaffilmark{1}, 
E.~Churazov\altaffilmark{2,3},
C.~Jones\altaffilmark{1}
S.~Heinz\altaffilmark{4},
R.~Kraft\altaffilmark{1},
A.~Vikhlinin\altaffilmark{1,3}
}

\altaffiltext{1}{Smithsonian Astrophysical Observatory,
Harvard-Smithsonian Center for Astrophysics, 60 Garden St., Cambridge,
MA 02138; wrf@cfa.harvard.edu}

\altaffiltext{2}{MPI f\"{u}r Astrophysik, Karl-Schwarzschild-Strasse
1, 85740
Garching, Germany}

\altaffiltext{3}{Space Research Institute (IKI), Profsoyuznaya 84/32,
Moscow 117810, Russia}

\altaffiltext{4}{University of Wisconsin, Madison, Wisconsin}

\begin{abstract}

M87, the active galaxy at the center of the Virgo cluster, is ideal
for studying the interaction of a supermassive black hole (SMBH) with
a  hot, gas-rich environment.  A deep Chandra observation of M87
  exhibits an approximately circular shock front (13 kpc radius, in
  projection) driven by the expansion of the central cavity (filled by
  the SMBH with relativistic radio-emitting plasma) with projected
  radius ~$\sim$1.9 kpc. We combine constraints from X-ray and radio
  observations of M87 with a shock model to derive the
properties of the outburst that created the 13~kpc shock.  Principal
constraints for the model are  1) the measured Mach number
  ($M$$\sim$1.2), 2) the radius of the 13~kpc shock, and 3) the observed
  size of the central cavity/bubble (the radio-bright cocoon) that
serves as the piston to drive the shock. We find an outburst of
$\sim$5$\times$$10^{57}$~ergs that began about 12~Myr ago and lasted
$\sim$2~Myr matches all the constraints.  In this model, $\sim$22\% of
the energy is carried by the shock as it expands. The remaining
$\sim$80\% of the outburst energy is available to heat the core
gas. More than half the total outburst energy initially goes into the
enthalpy of the central bubble, the radio cocoon. As the buoyant
bubble rises, much of its energy is transferred to the ambient thermal
gas. For an outburst repetition rate of about 12~Myrs (the age of the
outburst), 80\% of the outburst energy is sufficient to balance the
radiative cooling.

\end{abstract}


\section{The Outburst Chronicle of M87's Supermassive Black Hole}

The cavities and shocks observed in cluster, group, and galaxy images
of hot gas-rich systems chronicle the mechanical energy release, as
distinct from the radiated emission, from supermassive black holes
(SMBH) accreting at levels well below the Eddington mass accretion
rate ($\dot{M}_{\rm Edd}=4\pi G m_p M_{SMBH}/ \eta c \sigma_T$,
$M_{\rm SMBH}$ is the SMBH mass, $G$ is the gravitational constant,
$m_p$ is the proton mass, $c$ is the speed of light, $\sigma_T$ is the
Thomson electron scattering cross section, and $\eta\approx$10\%). For
present epoch SMBHs in  hot, gas-rich systems, the mechanical power
dominates the radiated power (e.g., Churazov et al. 2000, 2005, Fabian
et al. 2003, McNamara et al. 2005, Allen
et al. 2006, Fabian 2012). The best, and often
the only, way to derive the dominant energy release from the SMBH is
through the effects of the SMBH on the surrounding hot atmosphere.
The Eddington luminosity is given
as $L_{\rm Edd} = 1.3\times10^{47} (M_{SMBH}/10^9)$ \ergssec.
With an SMBH mass of
{$3-6\times10^9\>$\msun~ (Harms et al. 1994, Ford et al. 1994,
  Macchetto et al. 1997, Gebhardt et al. 2011, Walsh et al. 2013), the
  Eddington luminosity of M87's SMBH is $L_{\rm Edd} \sim 4-8 \times
  10^{47}$\ergssec.   The currently observed bolometric radiative
    luminosity $L_{rad}$ of the central AGN is
    $L_{rad} \approx3\times10^{42}$~\ergss (e.g., Prieto et al. 2016). This
    $L_{rad}$ is about five orders of magnitude lower than the
    Eddington limit for M87's mass, firmly placing the object into the
    category of low power AGNs. At the same time, the typical
    estimates of the jet mechanical power $L_{jet}$ of the source are
    consistently higher, $\sim10^{44}$~\ergss (e.g., Bicknell \&
    Begelman 1996, Owen, Eilek and Kassim, 2000, Stawarz et al.
    2006), implying that $L_{rad}/L_{jet}\sim0.03$ or lower. All these properties
    suggest that we are dealing with a variant of a hot, radiatively
    inefficient flow (e.g., Ichimaru 1977, Rees et al. 1982, Narayan \& Yi
    1994, Blandford \& Begelman 1999, Yuan \& Narayan 2014). M87's
    spectral energy distribution also supports this conclusion
    (Reynolds et al. 1996; Di
    Matteo et al. 2003; Yuan et al. 2009; Moscibrodzka et al. 2016).

\begin{figure*} [tb]
  \centerline{\includegraphics[width=0.99\linewidth]{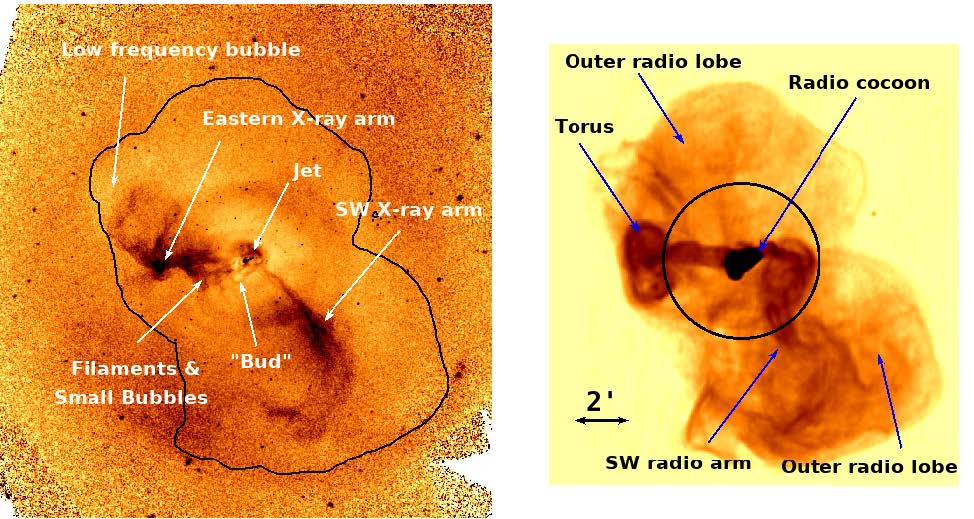}}
  \caption{{\bf(a - left)} The Chandra and 90~cm VLA (right) images,
    matched in scale, document the major outbursts from M87 (for
    details see Forman et al. 2007 and Owen et al. 2000). The Chandra
    image is a broad band image (0.5-2.5 keV) divided by the average
    radial profile to show faint surface brightness features.  A
    contour of the faintest surface brightness regions of the radio
    emission (0.15 mJy per $1.5''$ sq. pixel) is shown in the X-ray
    image. At the very core, the X-ray image shows the M87 jet
    (extending $20''$ to the NW) which is filling the central cavity
    with relativistic plasma clearly seen as the very dark (saturated)
    region in the radio image.  {\bf(b - right)} The VLA image shows a
    pair of arms extending up to $5'$ to the east and southwest. The
    eastern arm appears as a torus atop a stem (a ``mushroom cloud'')
    and represents a buoyant bubble of plasma that has risen about
    20~kpc over the past 40-70~Myrs (Owen et al. 2000, Churazov
      et al. 2001).  Only a twisted filamentary arm remains of the
    corresponding plasma bubble to the SW.  X-ray filamentary arms of
    cool gas, uplifted by the buoyant plasma bubbles, are seen in the
    Chandra image. On the largest scales (extending to almost 40~kpc,
    two faint disk-like radio features are probably the remnants of
    the oldest outbursts from M87 (of order $\sim100$ Myrs old;
    labeled as ``outer radio lobe''). The
    X-ray image shows a brightness enhancement surrounding the large
    radio structures that is most clearly visible to the south.  The
    X-ray image shows a shock at a radius of $2.8'$ (13 kpc; see also
    Fig.~\ref{fig:shock-image}), which is represented in the radio
    image as a black circle, that was produced by a prominent outburst
    approximately 12~Myr ago. }
\label{fig:xray-radio}
\label{fig:soft-radio}
\end{figure*}

\begin{figure*}
  \centerline{\includegraphics[width=0.99\linewidth]{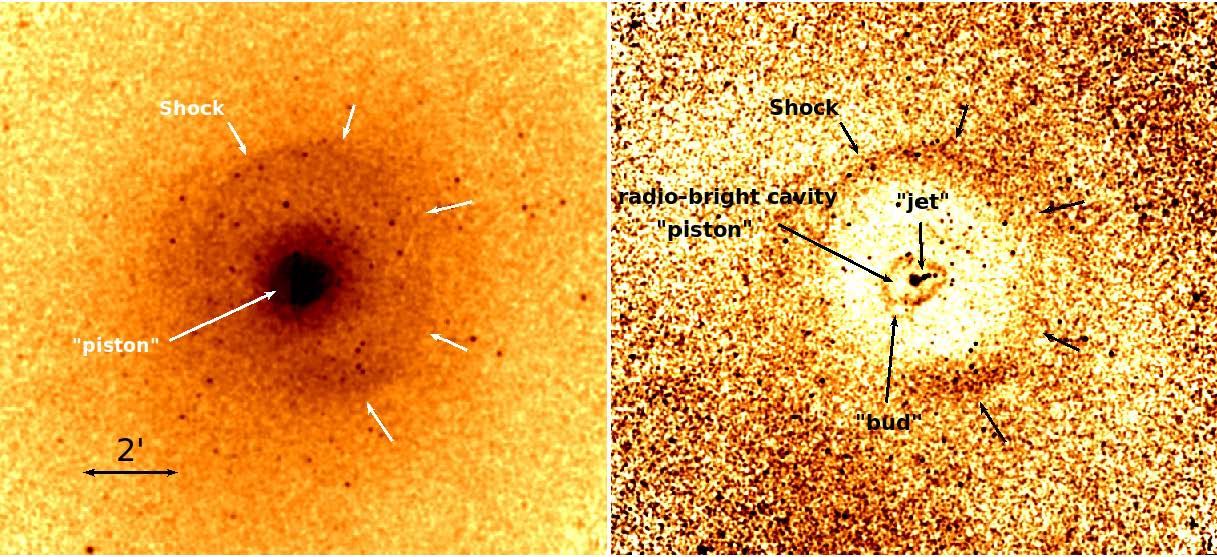}}
  \caption{Two renderings of the hard band (3.5-7.5 keV) Chandra image
    of M87.  As described in Forman et al. (2007;  see also
      Churazov et al. 2016), the hard energy band is approximately the
    square of the pressure projected on the sky for gas temperatures
    of 1 - 3 keV. Images in this band show direct evidence for
    outbursts as over-pressured regions. The cool filamentary arms, so
    prominent in the softer band (Fig.~\ref{fig:xray-radio} (left
    panel), are not seen in the hard band.  {\bf (a - left)} The
    hard-band image with background subtracted and corrected for
    vignetting.  {\bf (b - right)} The image on the right shows
    the data divided by the average radial profile to ``flatten'' the
    field and enhance features in the bright core as well as showing
    the low surface brightness outskirts.  The two panels show two
    clear outbursts -- the 13 kpc ($2.8'$) primary shock, and the
      central, over-pressured cocoon with the X-ray bright rim,
    initially inflated when the current 13 kpc shock began and now
    re-pressurized by the current outburst.}
\label{fig:hardimage}
\label{fig:shock-image}
\end{figure*}

X-ray and radio observations of M87 chronicle AGN outbursts over the
past 150~Myr.  The VLA radio observations from Owen, Eilek \& Kassim
(2000; see also de Gasperin et al. 2012 observations with LOFAR) show
evidence for the oldest outbursts  (see
  Fig.~\ref{fig:xray-radio}b).  The two filamented lobes lying NE and
  SW of the M87 nucleus have ages of $\sim100-150$~Myr. An eastern
  ``mushroom cloud'' with stem and torus and a filamentary
  southwestern arm (Fig.~\ref{fig:xray-radio}b) have estimated ages of
  40-70~Myr. X-ray filaments of cool gas ($\sim1$ keV) are seen
  coincident with these radio structures (compare
  Fig.~\ref{fig:xray-radio}a, b).  In addition, there are several less
  prominent features including 1) a bubble that is separating from the
  central cocoon (the ``bud'') seen in both X-ray and radio images (see
  Fig.~\ref{fig:xray-radio}b and Fig.~\ref{fig:shock-image}b), 2) a
  possible weak shock at about 5 kpc (about $10^6$ years old;
  see Fig.~\ref{fig:surbri}), 3) a series of filamentary structures extending to the
  east that are likely the remnants of small bubbles (see
  Fig.~\ref{fig:xray-radio}a); 4) a large cavity/bubble to the east
  (beyond the radio torus labeled as ``low-frequency bubble'' in
  Fig.~\ref{fig:xray-radio}a\footnote{This large cavity/bubble is
    very clearly detected in the LOFAR images just to the north of the
    torus, see Figs. 7 and 8 in de Gasperin et al. 2012)}; and 5) gas
sloshing cold fronts at large radii (33 kpc and 90 kpc; see Simionescu
et al. 2010 for a detailed discussion).


Recent major outbursts, in the past 20~Myr, are seen in a
combination of X-ray and radio imaging and are the focus of the
present paper.  The key features of these outbursts include:

\clearpage

\begin{itemize}
\item a classical shock at 13~kpc ($2.8'$) from the center of M87,
  seen in X-rays as a nearly complete azimuthal ring
  (Fig.~\ref{fig:hardimage} and Fig.~\ref{fig:surbri}).  This was the
  first classical shock found in the hot gaseous atmosphere around a
  central cluster galaxy where both the gas density and gas
  temperature jumps at the shock could be accurately measured. As
  Forman et al. (2007; see also Forman et al. 2005, Million et
  al. 2010) showed, the density and temperature jumps are separately
  consistent with the Rankine-Hugoniot shock jump conditions (Rankine
  1870, Hugoniot 1887) for a shock with a Mach number $M\sim1.2$.  The
  age of the outburst giving rise to the shock is about $12\times10^6$
  years.
\item the plasma-filled, radio-bright cocoon seen as an elongated
  bubble in the hard X-ray image (diameter $\sim40''$) that served as
  the piston to drive the 13~kpc shock and is, most likely, now being
  re-energized by  the present, ongoing outburst (see
    \ref{section:two} and 
    Figs.~\ref{fig:xray-radio}b, \ref{fig:hardimage}b,
    \ref{fig:cavity-size-cylinder}; also Hines et al. 1989).
\item the prominent jet, observed over a very broad wavelength range,
  flaring knots, and variable gamma-ray emission (Hines et al. 1989;
  Owen et al. 2000; Marshall et al. 2002; Harris et al. 2003, 2006;
  Shi et al. 2007; Forman et al. 2007; Abdo et al. 2009; Acciari et
  al. 2010).
\end{itemize}

\begin{figure} [thb] 
  \centerline{\includegraphics[width=0.950\linewidth]{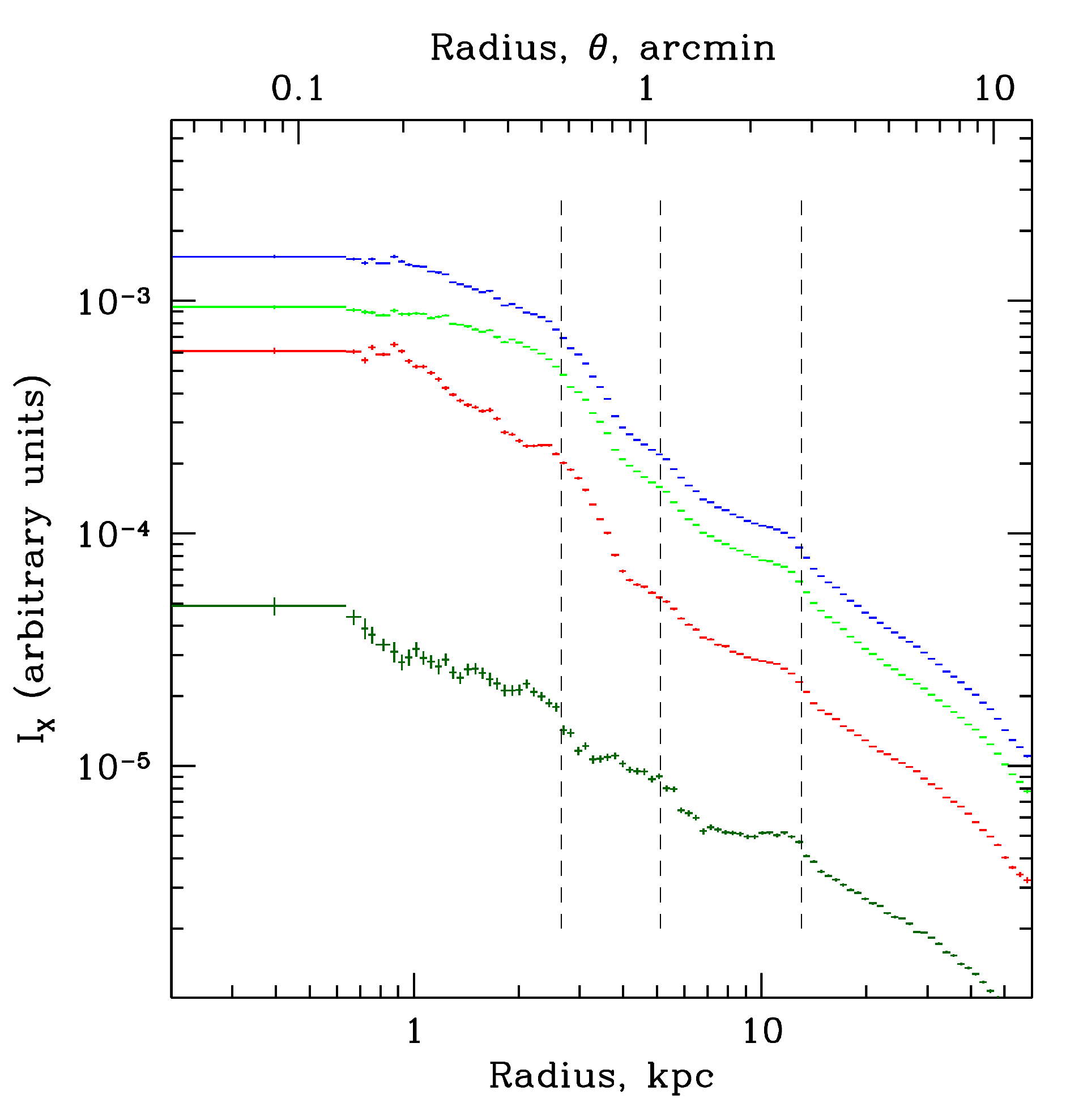}}
  \caption{Surface brightness profiles in four energy
    bands: broad (0.5-3.5 keV), medium (1.0-3.5 keV), soft (0.5-1.0
    keV), and hard (3.5-7.5 keV) from top-most to bottom-most. The
    surface brightness profiles are extracted from a 90\dg~ azimuth
    centered on north with  point sources excluded and corrected for
    vignetting and exposure. The three dashed vertical lines indicate
    the locations of features seen in the pressure maps
    (Fig.~\ref{fig:hardimage}). The inner most and outer most lines
    mark the
    strongest features and correspond to the current outburst that is
    re-inflating the central cavity and the 13~kpc shock. The 13~kpc
    shock is seen in all energy bands, while the central cavity is best seen
    in the hard band (lowest) surface brightness profile. A third
    weaker feature (possible shock) is seen at about $1'$ ($\sim 5$
    kpc; see also Forman
    et al. 2007, Million et al. 2010).}
\label{fig:surbri}
\end{figure}

The prominent 13~kpc
shock and its associated ``piston'' provide a unique opportunity to
investigate the energy balance between shock heating and heating from
buoyant bubbles inflated by AGN outbursts.  Fig.~\ref{fig:surbri}
shows the signature of outbursts in the observed surface
brightness profiles. Fig.~\ref{fig:deproj} shows the same signatures
in the deprojected  density and
temperature profiles. Fig.~\ref{fig:deproj}a is derived from the 360\dg
azimuthal average and provides the cleanestestimate of the mean gas
density properties, while Fig.~\ref{fig:deproj}b, a sector centered on
North, where the surface brightness profile is least affected by the
projection of cool filaments, provides the best estimates for the
shock parameters (see Forman et al. 2007 for the derivation of the
density and temperature jumps associated with the shock). This
``clean'' region in Fig.~\ref{fig:deproj}b shows the pronounced
enhancements in both temperature and density at the 13~kpc shock
($2.8'$) and at the outer edge of the piston at $\sim0.65'$
($\sim3$~kpc).

\begin{figure*} [thb] 
\centerline{\includegraphics[width=0.49\linewidth]{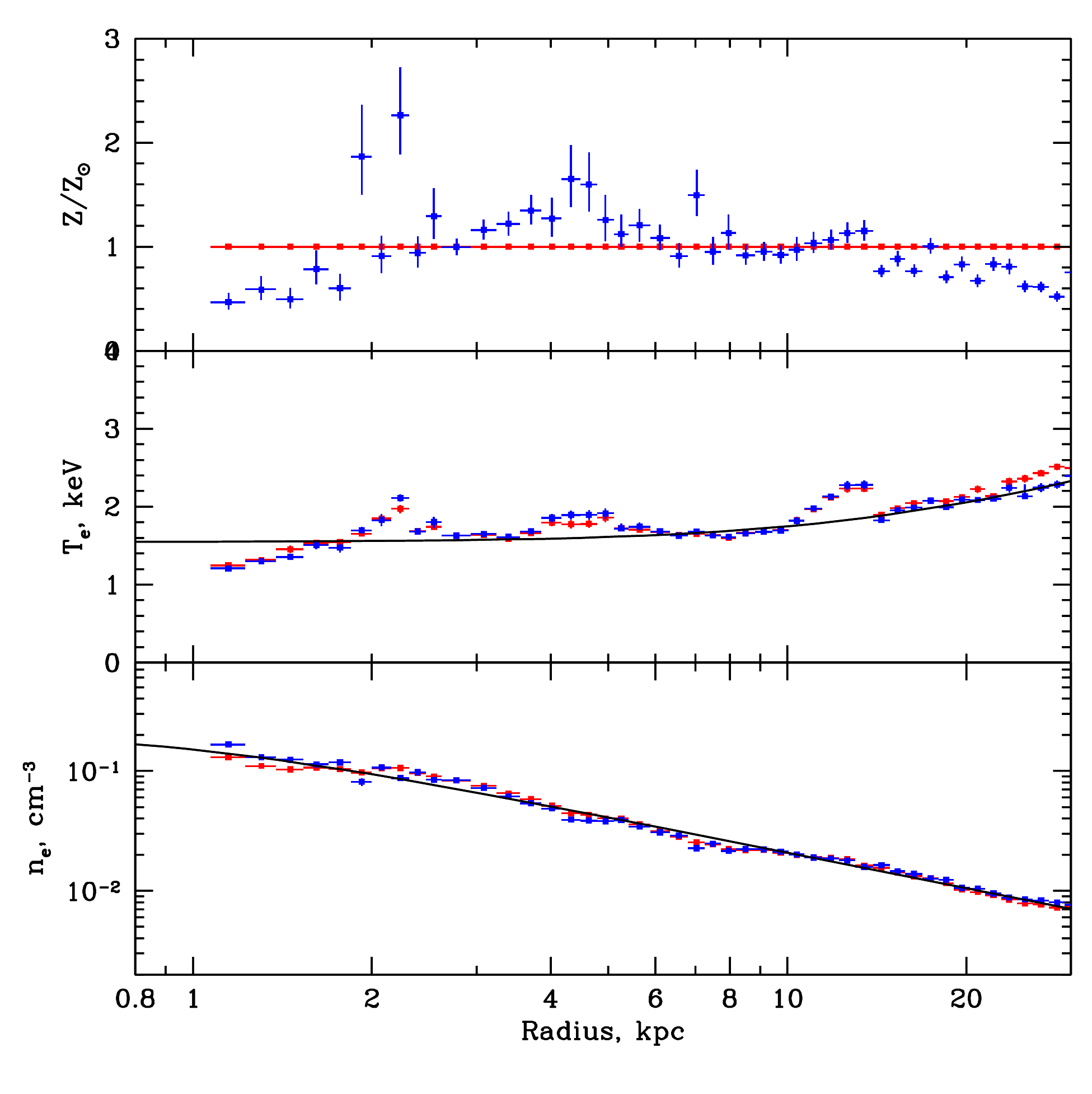}\includegraphics[width=0.490\linewidth]{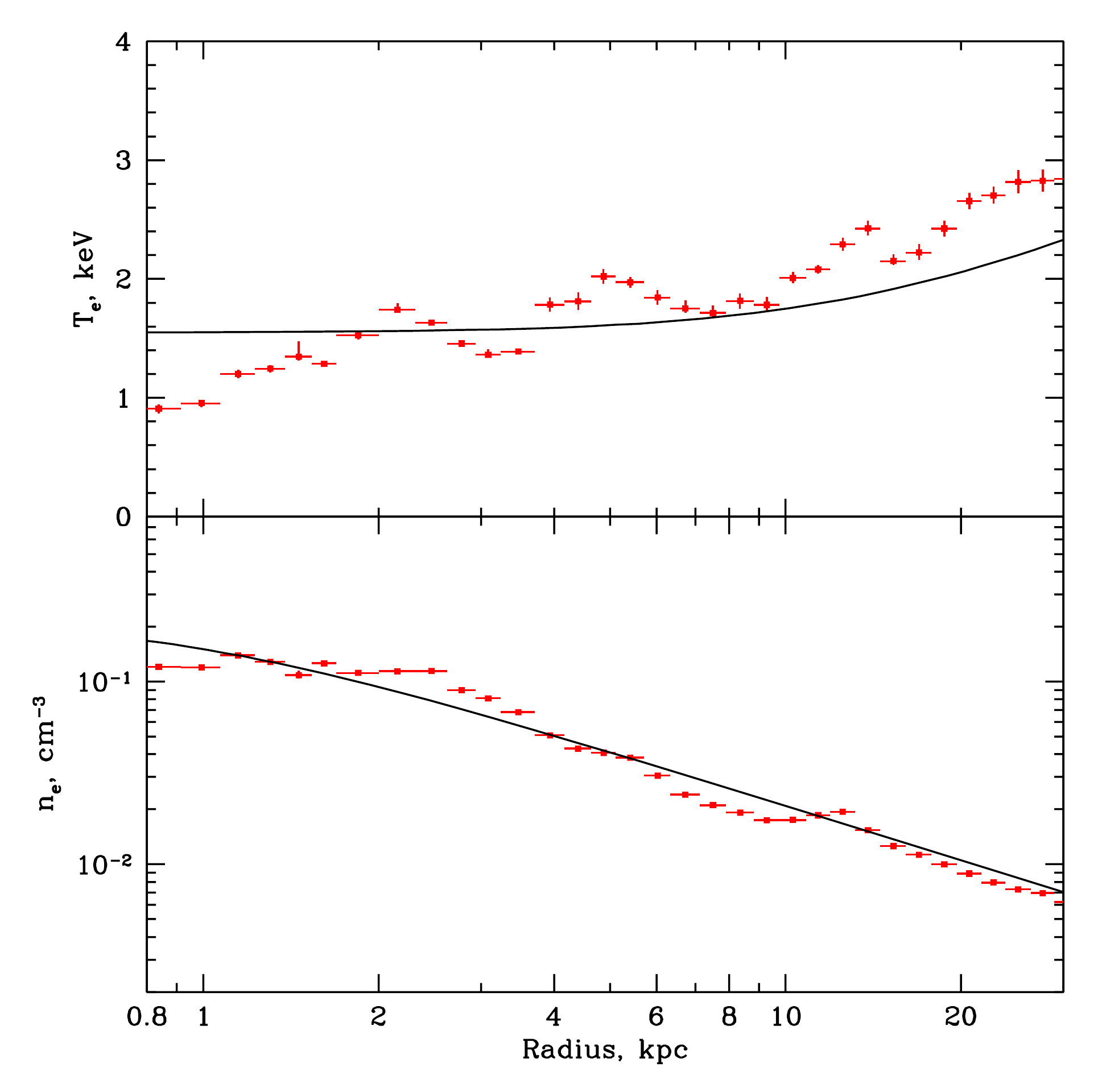}}
  \caption{{\bf (a - left)} The deprojected abundance, temperature and
    gas density profiles (red for abundance held fixed at solar and
    blue with variable abundance) derived from a full 360\dg~
    azimuthally averaged radial profile (excluding the prominent
      cool clump that lies almost exactly at the shock radius within
      the eastern arm and that distorts the average temperature
      profile. Excluding all other arm-like features makes little
      difference to the average profiles.).  The initial conditions
    are derived from fits to these data and are shown as the solid
    curves.  These curves also serve, in later discussions, as
      proxies for the data themselves, for comparison to the models.
    {\bf (b - right)} The deprojected gas density and temperature
    profiles derived from the 90\dg~ sector centered on North where
    the surface brightness profile is least disturbed by additional
    features, notably the soft X-ray arms. This ``clean'' region shows
    the pronounced enhancements in both temperature and density at the
    13~kpc shock ($2.8'$) and at the outer edge of the piston at
    $\sim0.65'$ ($\sim3$~kpc).  The solid curves are the fits to the
    complete 360\dg~ azimuthally averaged profile that serve as the
    initial conditions.
Spectral fits were done using
  an apec model and the deprojection procedure described in Churazov
  et al. (2008).}
\label{fig:fiducial-data}
\label{fig:deproj}
\end{figure*}

We investigate M87's recent outburst history by using a 1-D numerical
shock model to characterize the observed properties including the gas
temperature and density profiles. Because the outburst has occurred in
the cool atmosphere of M87, compared to hotter atmospheres in more
luminous clusters, we are able to derive the observable quantities of
the outburst in considerable detail (see Forman et al. 2007, Churazov
et al. 2008).  By combining a simple model with the high quality
observations of M87, we can determine the parameters of the outburst
and the energy partition between the shock and the cavity enthalpy and
thus help understand the different heating mechanisms required to
suppress strong cooling flows in hot atmospheres in galaxies, groups,
and clusters.

\subsection{Cavity Size}
\label{section:cavity-size}

One of the key constraints on the outburst model comes from the
size/volume of the central cavity produced as the relativistic plasma
from the jet displaces the hot X-ray emitting gas in the core of
M87. The appropriate size to be used is complicated by the fact that
the jet is double-sided and inclined to the plane of the sky. As a
result, the jet is probably producing two cavities that together make
an elongated structure rather than a single spherical cavity.

\begin{figure*} 
\centerline{\includegraphics[width=0.97\linewidth]{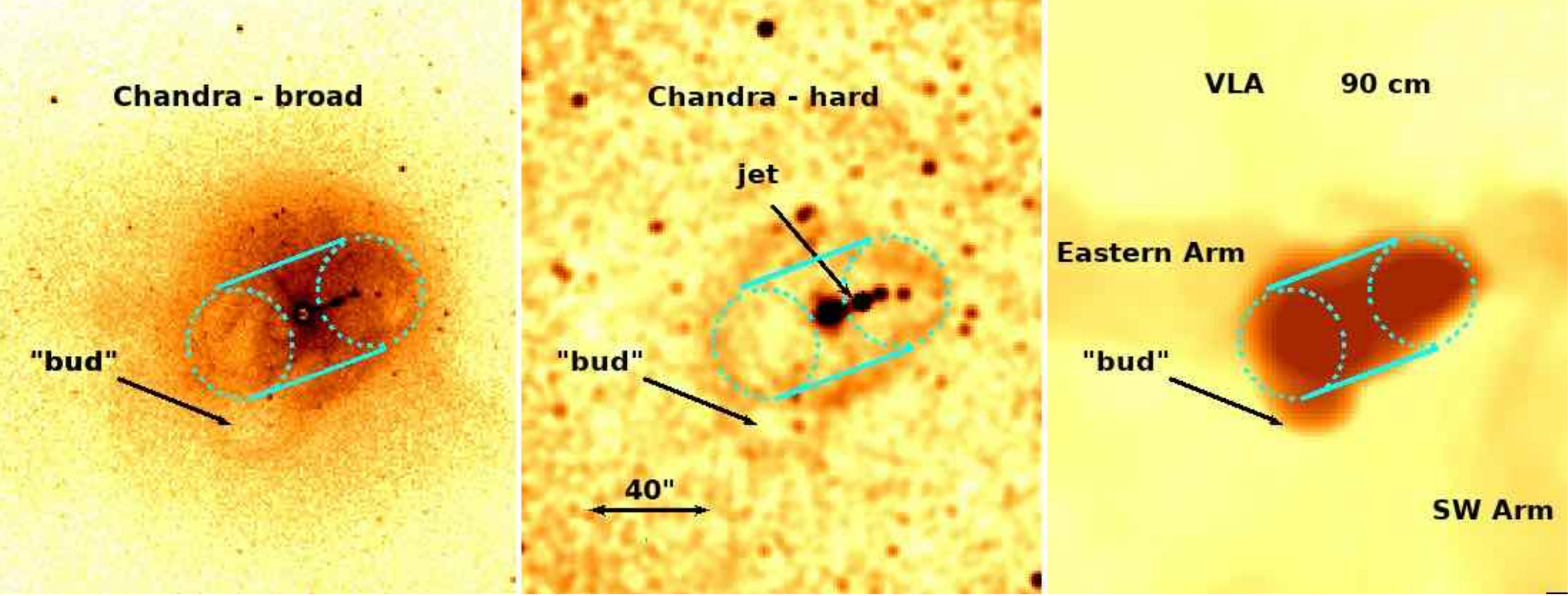}}
 \caption{Images of the M87 core with the central cavity
   approximated as a cylinder (front and rear surfaces as dashed
   circles and sides as solid lines) with various features labeled.  The three images of the core
   are: (left) Chandra broad band image (0.5-2.5 keV) divided by the
   average radial profile, (center) Chandra hard band image (3.5-7.5
   keV) divided by the average radial profile, and (right) 90~cm VLA
   image of M87. The eastern and
   southwestern (SW) arms are very faint compared to the highly
   overexposed cocoon.  }
\label{fig:cavity-image}
\label{fig:cavity-size-cylinder}
\end{figure*}

For a proper comparison with the predictions of the 1D model, it is
important to estimate the bubble volume in 3D, since the $P V$ work
required to displace the X-ray emitting gas is the most direct proxy
for the total energetics of the outburst in the model with ``gradual''
energy release (see Section 2.5 below). To this end, we have
approximated the cavity as an inclined cylinder, co-aligned with the
jet axis (Fig.~\ref{fig:cavity-size-cylinder}). Projected on the sky,
the cylinder consists of a circular cross section with radius $0.3'$
and height $1.1'$. Inclination angles for the M87 jet range from 10\dg
-- 20\dg (e.g., Biretta, Sparks \& Macchetto 1999, Wang \& Zhou 2009).
Taking the volume as the geometric mean from the two extreme
inclinations and converting this to a sphere gives a spherical volume
with a radius of $\sim3$~kpc (equivalent to $0.65'$).  For our 1D model,
we use this value in our calculations.

The X-ray cavity size matches that of the radio cocoon/bubble
(Fig.~\ref{fig:cavity-image}) and we typically refer to the ``cavity'' in the
discussion of the model.

\section{Simulations of the M87 13 kpc Shock}

Our simulations are carried out in the context of a simple outburst
model that captures the key physics.  The radio plasma, ejected from
the supermassive black hole by the jet, inflates a central cavity,
seen as lobes or a cocoon in M87 radio maps
(Fig.~\ref{fig:cavity-size-cylinder}, right panel).  The inner radio lobes
act as a piston that displaces the X-ray emitting plasma.  Our results
are uncertain due to projection effects arising from the unknown
geometry and since we do not know the precise initial conditions of
the M87 atmosphere, prior to these SMBH outbursts.  Also, we neglect
possible effects of diffusive processes on the weak shock (cf. Fabian
et al. 2006). However, as we show, the qualitative features of
the density and temperature profiles provide a robust characterization
of the outbursts.

\subsection{Numerical Modeling Details}

We have performed a sequence of 1D Lagrangian numerical simulations of
a shock propagating into the M87 atmosphere where we vary the energy
deposited by the outburst and the timescale over which the energy is
injected by the central AGN.  The M87 atmosphere is assumed to lie in a
static gravitational potential, $\phi (r)$, such that the 
observed gas density and gas temperature distributions (see
section~\ref{section:initial}) are in
hydrostatic equilibrium.   We assume, for the initial
  conditions, that the present M87 gas density and temperature are
  close to those prior to the outburst, i.e., M87's atmosphere is in a
  ``steady state'' with repeated outbursts that are not unusually
  violent.

\begin{figure} [thb] 
  \centerline{\includegraphics[width=0.950\linewidth]{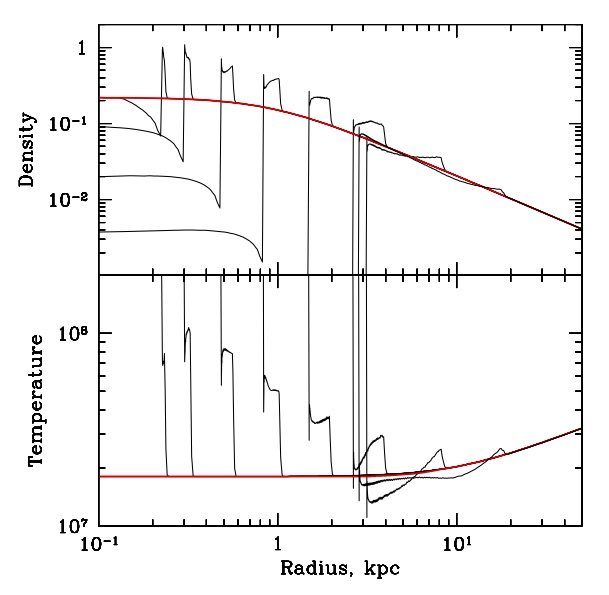}}
  \caption{The gas density and temperature distributions of the
    fiducial model as a function of time.  The shock is initially
    strong with both the gas density and gas temperature jumps
    decaying with time.  The eight models shown are snapshots taken at
    0.023, 0.061, 0.16, 0.41, 1.07, 2.77, 7.18, $18.6\times10^6$ years
    after the initial outburst.  The particular model shown, with an
    outburst energy of $5.5\times10^{57}$ ergs and a duration of
    $2.2\times10^6$ yrs, matches 1) the best fit Mach number ($M=1.2$)
    at the 13 kpc radius of the observed shock and 2) the estimated
    central cavity (piston) radius of $\sim3$~kpc. Since this model captures
    the key parameters of the outburst, it is referred to as the fiducial model.  The
    initial conditions are shown as a solid red line (given in
    equations 1 and 2).  The temperature
    interior to the piston reflects that for the mixture of very hot
    relativistic plasma that mixes with the small quantity of thermal
    gas present in the inner pixels of the model when the outburst
    begins.}
\label{fig:fiducial-evolution}
\end{figure}

We assume that an outburst from a SMBH deposits an energy $E_0$
uniformly over a time interval $\Delta t$.  In the inner cells
interior to the boundary of the  piston (initially 0.2~kpc), the energy is deposited as a
power law in radius to mimic the deposition of energy as a jet fills
the central cavity (see Xiang et al. 2009 for additional details).
For all gas components, we assumed in the actual calculations that
$\gamma=5/3$.  For a cavity of radius $R$, pressure $P$ (in
pressure equilibrium with the ambient gas) and
volume $V$, the minimum total energy required to inflate the cavity is
$E_{tot}=\gamma/(\gamma-1) P V$.  Since the component interior to the
piston is at least partially
a relativistic plasma, the appropriate $\gamma$ may be smaller and the
input energy larger.   For
$\gamma = 4/3$ and subsonic expansion, $E_{tot}$ would be 60\% larger
than for $\gamma = 5/3$.  We discuss the implications of different
values for $\gamma$ in section~\ref{sec:gamma}.

\begin{figure} 
  \centerline{\includegraphics[width=0.95\linewidth]{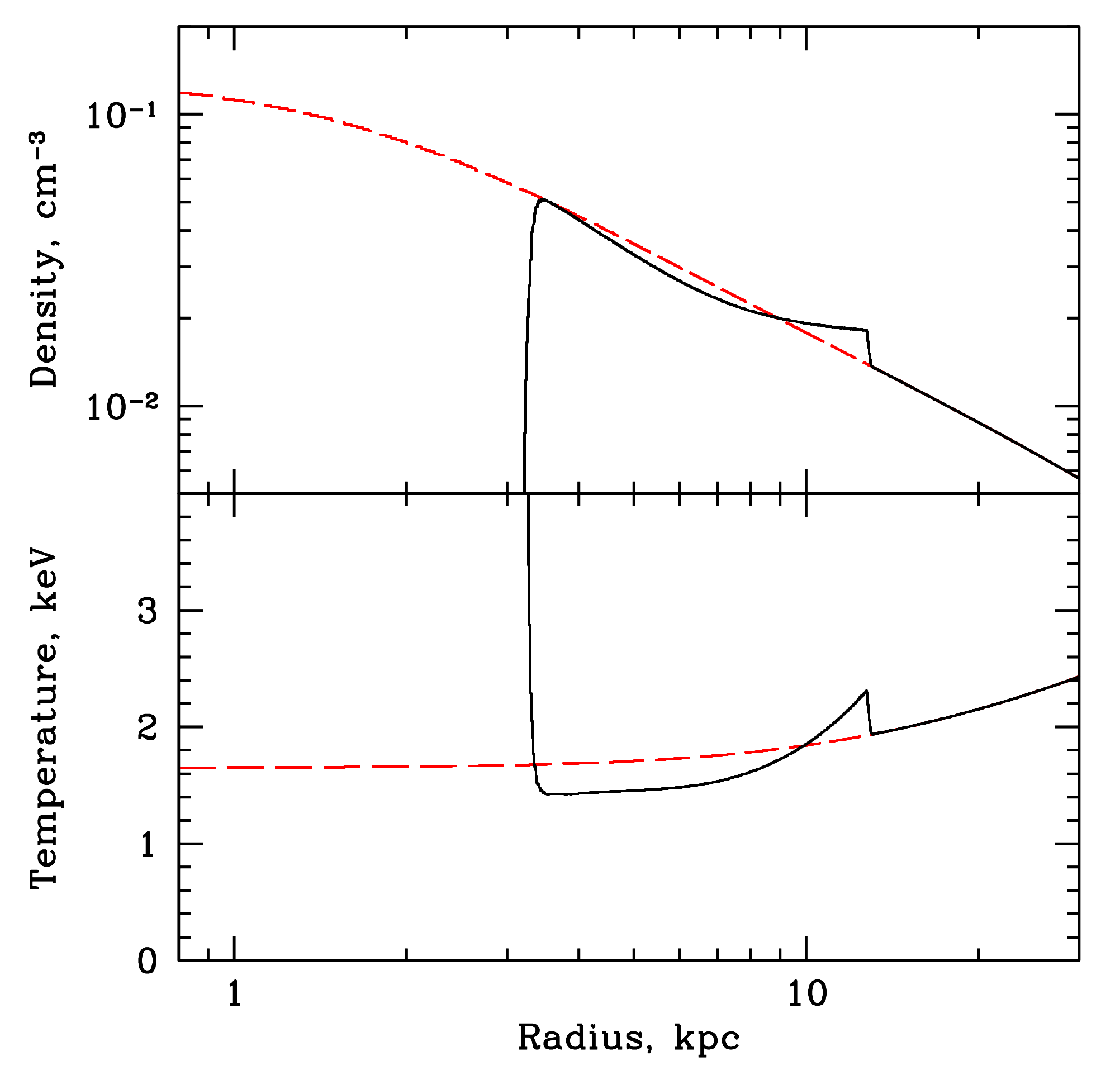}}
 \caption{Initial radial profiles of density and temperature (the
   initial conditions) as modified by the outburst. The dashed red
   lines correspond to the initial conditions.
   The black solid lines show the density and temperature profiles
   that characterize the ``fiducial 1D model'' with total energy
   release $5.5\times 10^{57}$ ergs and outburst duration $2x10^6$ yr,
   when the shock front is $\sim13$ kpc from the center of the
   cluster. For the fiducial model (and for ``long'' outburst models
   in general), downstream from the shock, the gas temperature is
   lower than the initial temperature of the gas at the same
   radius (for reasons described in the text). 
}
\label{fig:fiducial-model}
\end{figure}

\subsection{Initial Conditions}
\label{section:initial}

The initial conditions of the hot gas surrounding M87
are a fundamental input to the
model. Despite the high quality Chandra X-ray observations, the
conditions of the atmosphere surrounding M87, as they appeared more
than 10 Myrs ago,  prior to the outburst are uncertain, since the gas surrounding M87 has
experienced a variety of outbursts (and possibly even  small  mergers
and the associated ``gas sloshing'').  However, as a dynamically old
system with an old stellar population, we assume that the atmosphere
around M87 is in quasi-equilibrium and has not undergone any dramatic
changes in recent epochs. If, as seems likely, the SMBH in M87 is able
to maintain a quasi-equilibrium between heating and radiative cooling,
then the present  is a ``fair''  match to the conditions that were
present at the time of the outburst.  Therefore, for the region
interior to $6'$ ($\sim 30$ kpc), we use the observed gas density and
temperature distributions to derive the ``unperturbed'' gas density and
gas temperature profiles that are fit to the deprojected data with the
simple analytic functions:
\begin{eqnarray}
n_e(r) &= 0.22  (1+(r/r_c)^2)^{-3\beta/2} \\
\label{eq:fiducial1}
kT (r) &= 1.55 (1 +(r/r_T)^2)^{0.18} 
\label{eq:fiducial2}
\end{eqnarray}
where $r_c= 0.2'$ (0.93 kpc), $\beta=0.33$, and $r_T = 2.2'$ (10.2
kpc). These profiles, derived from the full 360\dg~ azimuthal average
profile are shown in Fig.~\ref{fig:deproj}a and provide the
initial baseline for the simulations. Fig.~\ref{fig:deproj}b
shows the initial conditions compared to the observations of the northern sector 
where the shock is most clearly seen.

\begin{figure*} 
  \centerline{\includegraphics[width=0.49\linewidth]{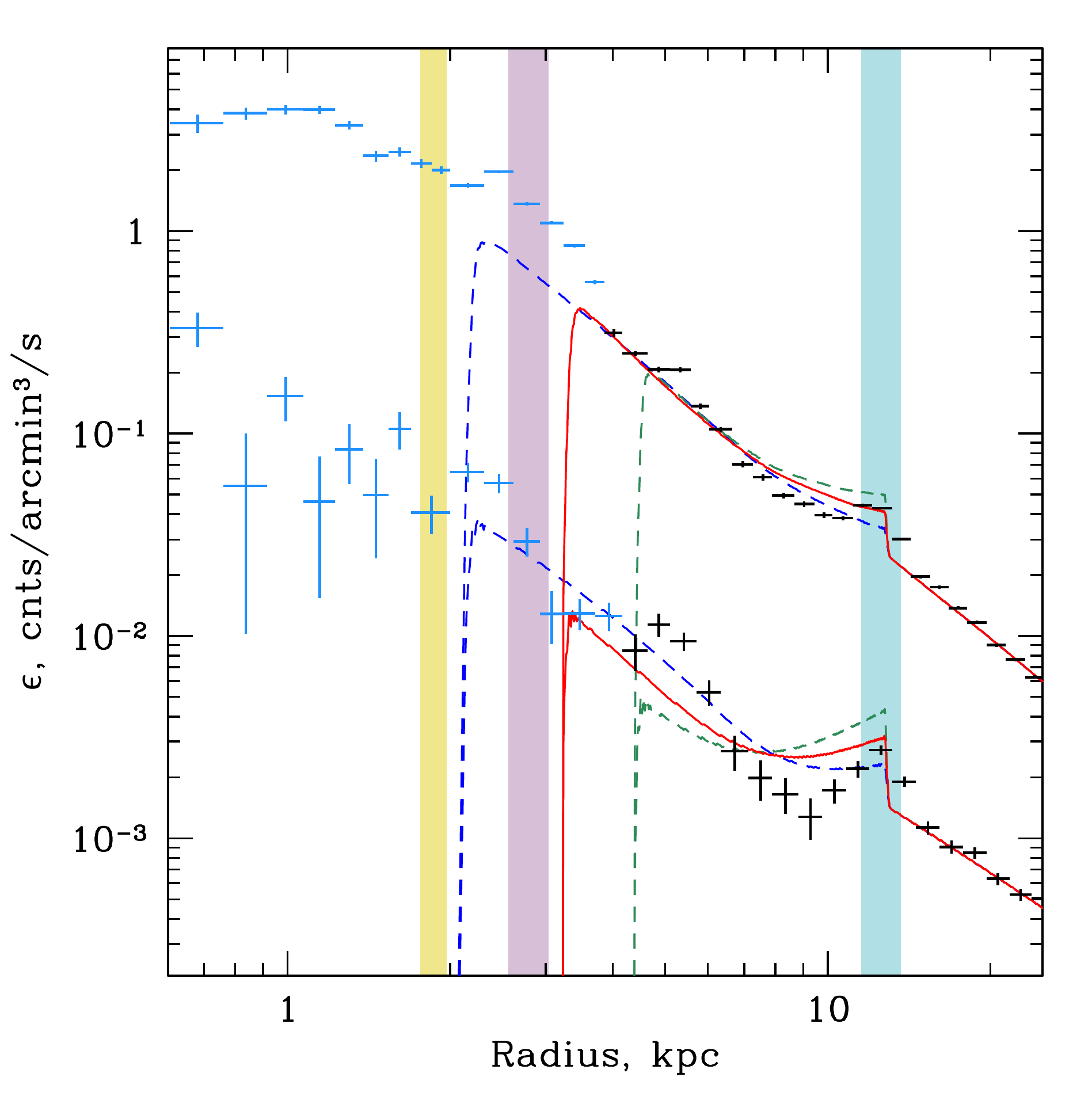} \includegraphics[width=0.49\linewidth]{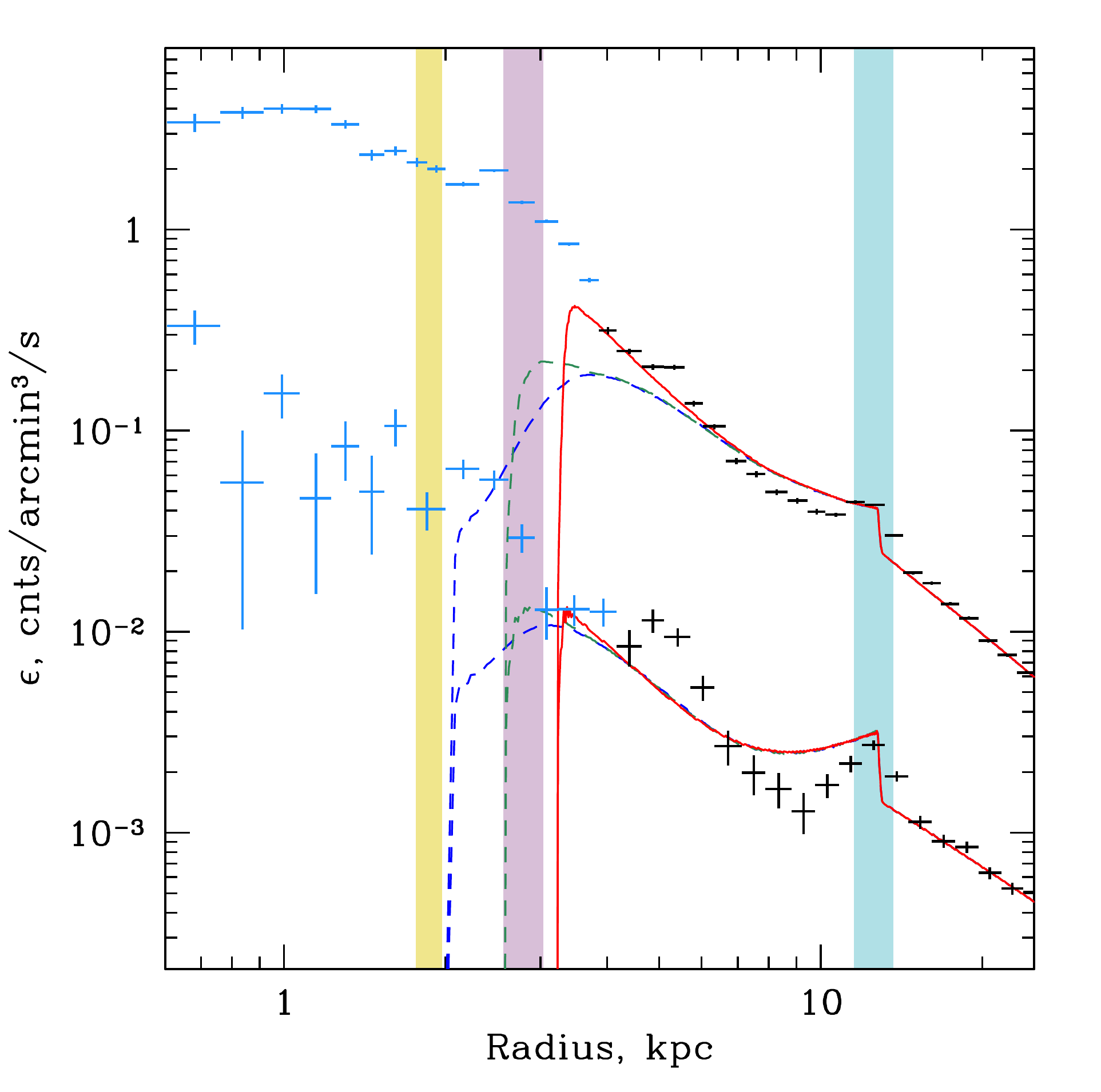}}
\caption{ Comparison of outburst models with the data (the radial
  sector from 45-135\dg) for M87. The deprojected emissivity in two
  energy bands (upper data points are from 0.5-3.5 keV; lower data
  points are from 3.5-7.5 keV) is compared to i) models with varying
  outburst energy and fixed outburst duration (left panel) and ii)
  models with varying outburst durations and fixed outburst energy
  (right panel).  The three vertical bands, from right to left,
  indicate the location of the shock (blue), ``effective'' cavity
  radius (estimated in section~\ref{section:cavity-size} after
  accounting for the line-of-sight projection) (magenta), and
  ``apparent'' cavity radius projected on the sky (yellow). The data
  points interior to the piston, within the region partially filled by
  the radio cocoon (radii less than about 3~kpc) are shown as light
  blue to indicate that they are dominated by systematic uncertainties
  including overlying complex structures, a highly uncertain
  deprojection (since the volume is partially filled with an uncertain
  amount of radio plasma), and should not be considered in
  comparisons to the model. The three models in the left panel have
  the same outburst duration, but different energies of 2.0, 5.5, 11
  $\times 10^{57}$~ergs (violet, red, green) respectively, leading to
  different amplitudes of the shock. The three models in the right
  panel have, on the contrary, the same outburst energy of
  $5.5\times10^{57}$ ergs, but different durations of 0.05, 0.56, and
  2.2 Myr (violet, green, red), respectively, leading to different
  sizes of the central cavity. Thus the jumps in density/temperature
  and the size of the cavity together can naturally constrain the
  parameters of the outburst. The red curves in both panels correspond
  to the fiducial model that reproduces the major observables.}
\label{fig:model-data}
\end{figure*}

\begin{figure*}
  \centerline{
    \includegraphics[height=0.49\linewidth]{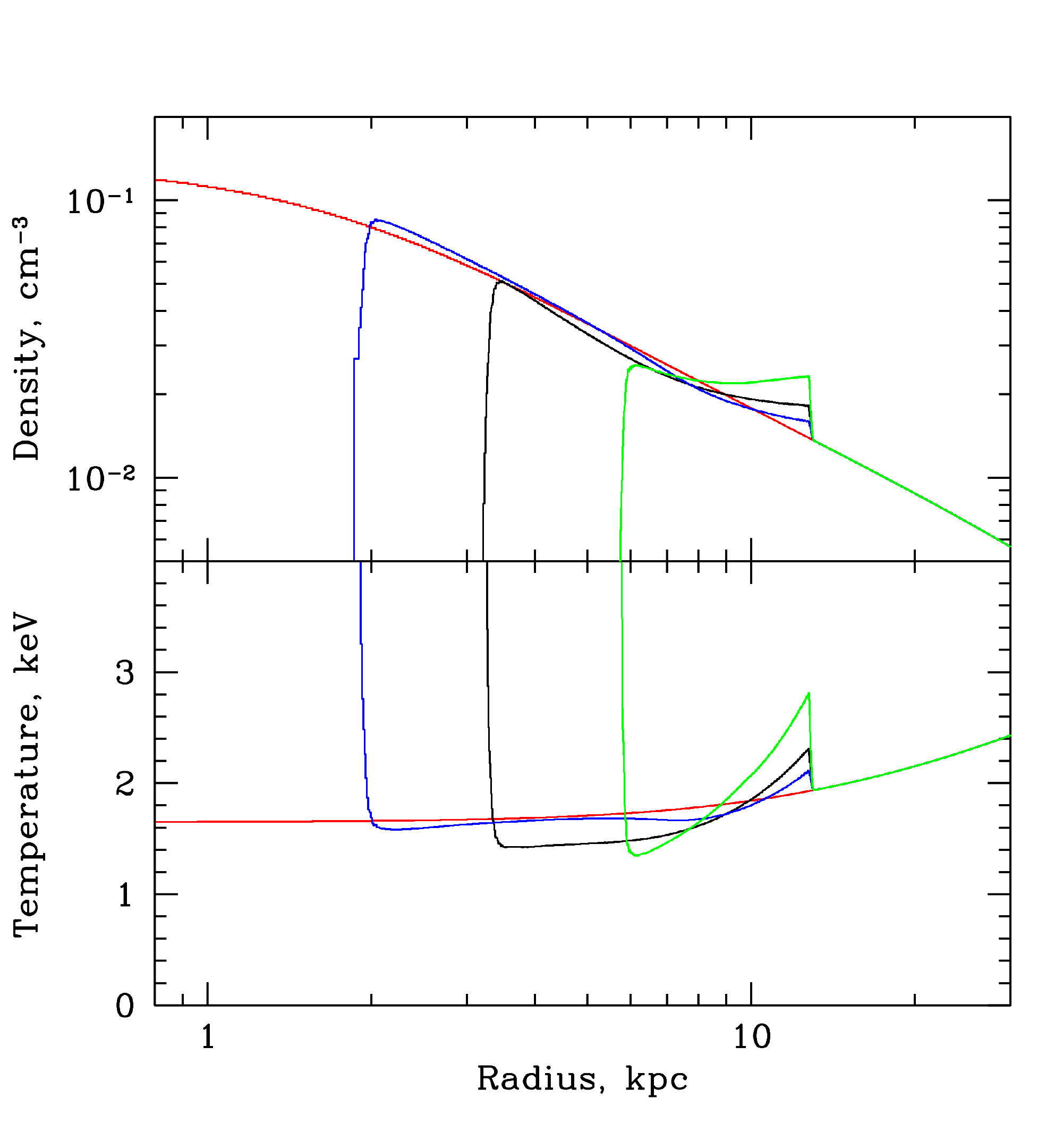}
    \includegraphics[height=0.49\linewidth]{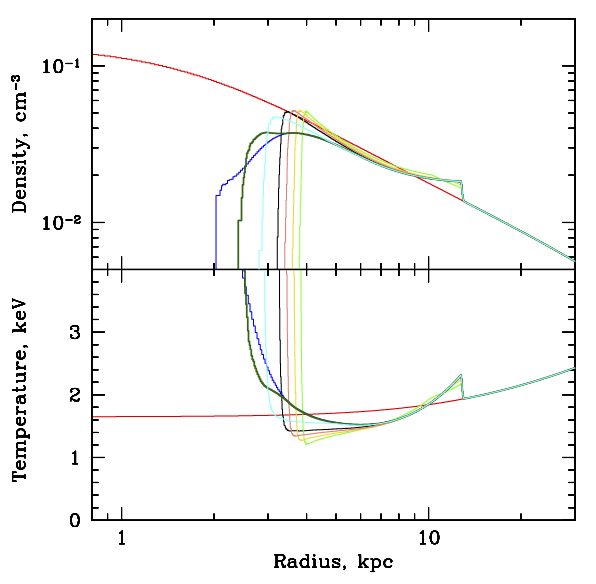}}
%
  \caption{{\bf (a-left)} For an outburst duration of $2.2\times10^6$
    yrs, the fiducial value, we show a series of models with outburst
    energies of 1.4, 5.5, $22 \times 10^{57}$ ergs when the shock lies
    at a radial distance of 13 kpc and hence having different
    ages).  As the models show, the outburst energy is the major
  contributor to the amplitude of the shock.  {\bf (b-right)} For an
  outburst energy $E = 5.5\times 10^{57}$ ergs, we show a set of
  models when the shock has reached a radius of 13 kpc for outburst
  durations of 0.1, 0.4, 1.1, 2.2, 3.1, 4.0, 4.4, $6.2 \times10^6$
  yrs.  The magnitude of the shock is independent of duration for
  durations less than about $4\times10^6$ yrs.  For longer duration
  events, the cavity is still significantly over-pressurized. As shown
  in Fig.~\ref{fig:grid}, the duration is constrained by the
  combination of shock strength and piston/cavity size.}
\label{fig:duration}
\label{fig:energy}
\vspace*{6in}
\end{figure*}

\subsection{A Shock in the Atmosphere of M87}

Applying our shock model to the initial conditions described above, we
can examine a typical outburst.  Our ``fiducial'' model with total
outburst energy and outburst duration $E_{tot}=5.5\times10^{57}$~ergs
and $\Delta t=2$~Myr has a temporal evolution shown in
Fig.~\ref{fig:fiducial-evolution}. This temporal evolution is
characteristic of all the models.  The initial shock weakens with
time, because of energy dissipation at the front at early phases, when
the shock is still strong, and undergoes pure spherical expansion at
later phases. In the last snapshots, the shock is expanding at Mach
$M=1.2$ with amplitudes, in both density and temperature, that match
the observations.  The expansion of the central cavity (the piston)
``stalls'' at the present observed piston radius of about 3~kpc.
  In fact, the inertia of the accelerated gas ahead of the piston
  carries it beyond the pressure equilibrium radius and the piston
  radius subsequently decreases slightly in the last time steps.  This
  effect also is seen in the 3D simulation that we used to confirm the
  validity of our 1D models (described in section~\ref{sec:3d}), but
  the effect is less pronounced.  The final configuration, as we show
  below, matches the observations and for this reason, the outburst
  with $E_0 = 5.5\times10^{57}$ ergs and $\Delta t = 2 \times 10^6$
  yrs is referred to as the fiducial model.}

   Fig.~\ref{fig:fiducial-model} shows the same fiducial model at
  the moment when the shock front reaches 13 kpc, corresponding to the
  observed shock radius.  For the fiducial model (and for ``long''
  outburst models in general), downstream from the shock, the gas
  temperature is lower than the initial temperature of the gas at the
  same radius.  This is due to a combination of two effects. First,
  the rarefaction region behind a shock is a generic feature of weak
  spherical shocks (as described by Zeldovich \& Razier 2002 and
  Landau \& Lifshitz 1959).  Second, in these models, the adiabatic
  expansion of the gas that is displaced from its initial location to
  lower pressure regions (larger radii) contributes to the temperature
  decrease. These features can be identified in many of the figures in
  this paper.

The lack of perfect spherical symmetry, the presence of cool
structures (arms), and the uncertainty in the initial conditions
complicate any detailed, quantitative comparison of the model and
data. However, a qualitative (``factor of 2'') comparison is
possible. Since the wedge to the North is less contaminated by cool
structures, except for the inner $45''$, we used the deprojected
emissivities in the 0.5-3.5 and 3.5-7.5 keV bands for comparison with
the model predictions (see Fig.~\ref{fig:model-data}). The emissivity
in these two energy bands was calculated using the predicted density and
temperature profiles assuming fixed solar metalicity. For the models
shown in Fig.~\ref{fig:model-data}, the fiducial model captures the
key parameters measured for the M87 outburst and matches the size of
the central cavity, the observed radius and strength of the shock (in
both density and temperature), and the emissivity outside the central
cavity. 

  As noted above, none of the 1D models provides a ``perfect fit''
  to the data over the entire radial range. This is especially true
  for the innermost part, where the 1D model predicts the complete
  evacuation of the gas as it is pushed away by a spherical piston. In
  a real cluster, the gas is expected to be evacuated only from
  regions occupied by the cavities (the radio plasma), while the
  thermal gas can still be present along other directions. This is why
  we will compare the size of the cavity predicted by the 1D simulations to
  that derived in section~\ref{section:cavity-size}, rather than
  directly comparing the predicted and observed profiles. For the
  shock front region, which is farther away from the center, the
  effects of asymmetry should be less severe and the direct comparison
  of the radial profiles is better justified.

\begin{figure*} [thb]
  \centerline{\includegraphics[width=0.95\linewidth]{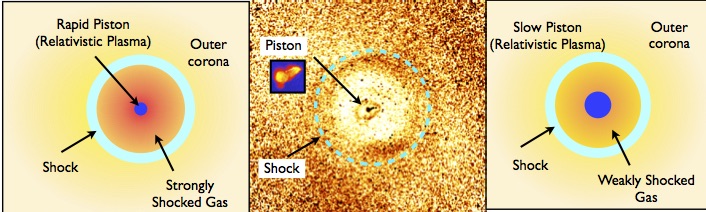}}
  \caption{Schematic
    of the two shock scenarios - a short and a longer
      outburst. The modeled gas density and gas temperature profiles
    are shown in Fig.~\ref{fig:slow-fast}. {\bf (a - left)} A
    powerful, short duration outburst ($\Delta t = 0.1 \times 10^6$
    yrs) drives a strong shock into the surrounding atmosphere. At the
    present time, the region interior to the shock (located at the
    observed 13 kpc radius) would enclose a central hot, 
      strongly shocked, low density atmosphere. {\bf (b - center)}
    X-ray image of M87 divided by the average radial profile to better
    show the central piston and the jet  with an inset image of
      the 6 cm radio emission that shows the piston that drove the M87
      outburst.  The dashed blue circle (labeled ``Shock'')
    indicates the outer edge of the shock which is seen as the bright
    ring of emission.  {\bf (c - right)} A longer duration outburst,
    ($\Delta t = 2.2\times10^6$ yrs;  the fiducial model)
    provides the same magnitude shock at 13 kpc, but only weakly
    shocked gas interior to the shock location and a larger central
    plasma-filled, piston. As discussed in the text, short duration
    outbursts are inconsistent with the observations.}
\label{fig:models-schematic}
\end{figure*}

\subsection{Effects of Outburst Energy and Outburst Duration}

To explore the range of allowed outburst parameters, we separately
investigate the effects of varying the outburst energy and outburst
duration.  These two parameters govern the final outburst
configuration.

For a given outburst duration, the outburst energy strongly affects
the amplitude of the shock.  Fig.~\ref{fig:energy}a shows the gas
density and temperature when the shock reaches 13~kpc, for outburst
energies of $1.4, 5.5, 22 \times 10^{57}$ ergs.  The choice of
outburst energy brackets the energy described above as the fiducial
value.   As Fig.~\ref{fig:energy}a shows, the amplitude of the shock
alone provides a direct diagnostic of the outburst energy.  Also, note that
the different values of the outburst energy yield different sizes
for the central piston -- larger energy outbursts drive stronger
shocks that reach 13 kpc in a shorter time and have larger central
cavities of relativistic plasma.

We also have investigated the effects of varying the outburst
duration.  Fig.~\ref{fig:duration}b shows the gas temperature and gas
density profiles for models with the outburst energy held fixed
at $E_0 = 5.5 \times 10^{57}$ ergs and with outburst durations ranging
from $0.1$ to  $6.2 \times 10^{6}$ yrs.  While the amplitude of the shock at 13
kpc varies only slightly, the size of the piston varies
dramatically. 
The models show the characteristic behavior of ``short'' and ``long''
duration outbursts.  As we show below, by matching the observations to
the models in more detail, we can estimate a quantitative value for
the outburst duration. Also, as Fig.~\ref{fig:duration}b shows, a ``short''
duration outburst produces a central region with $\sim2-3$~kpc
radius), starting just beyond the outer boundary of the piston, that
consists of hot, low density gas.  In contrast, the longer duration,
initially weaker shocks, with the same total outburst energy, are
bounded by cool shells and have no extended hot, shocked region (see
also Brighenti \& Mathews 2002).

Thus, the combination of Fig.~~\ref{fig:energy}a and
\ref{fig:duration}b shows that the outburst energy is determined
(primarily) by the magnitude of the jumps.

\begin{figure*} [t]
\centerline{
  \includegraphics[width=0.49\linewidth]{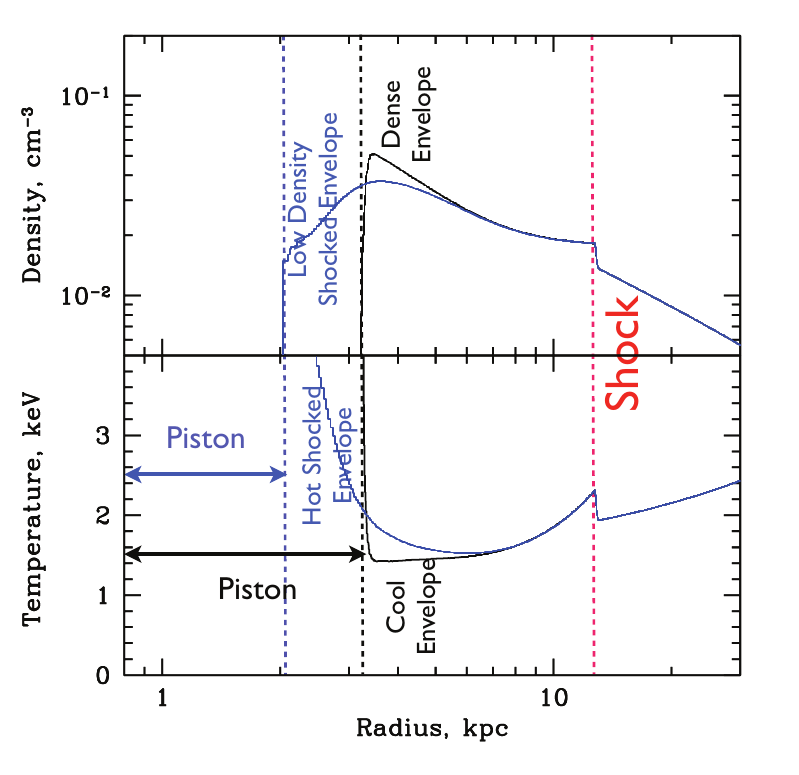}
  \includegraphics[width=0.49\linewidth]{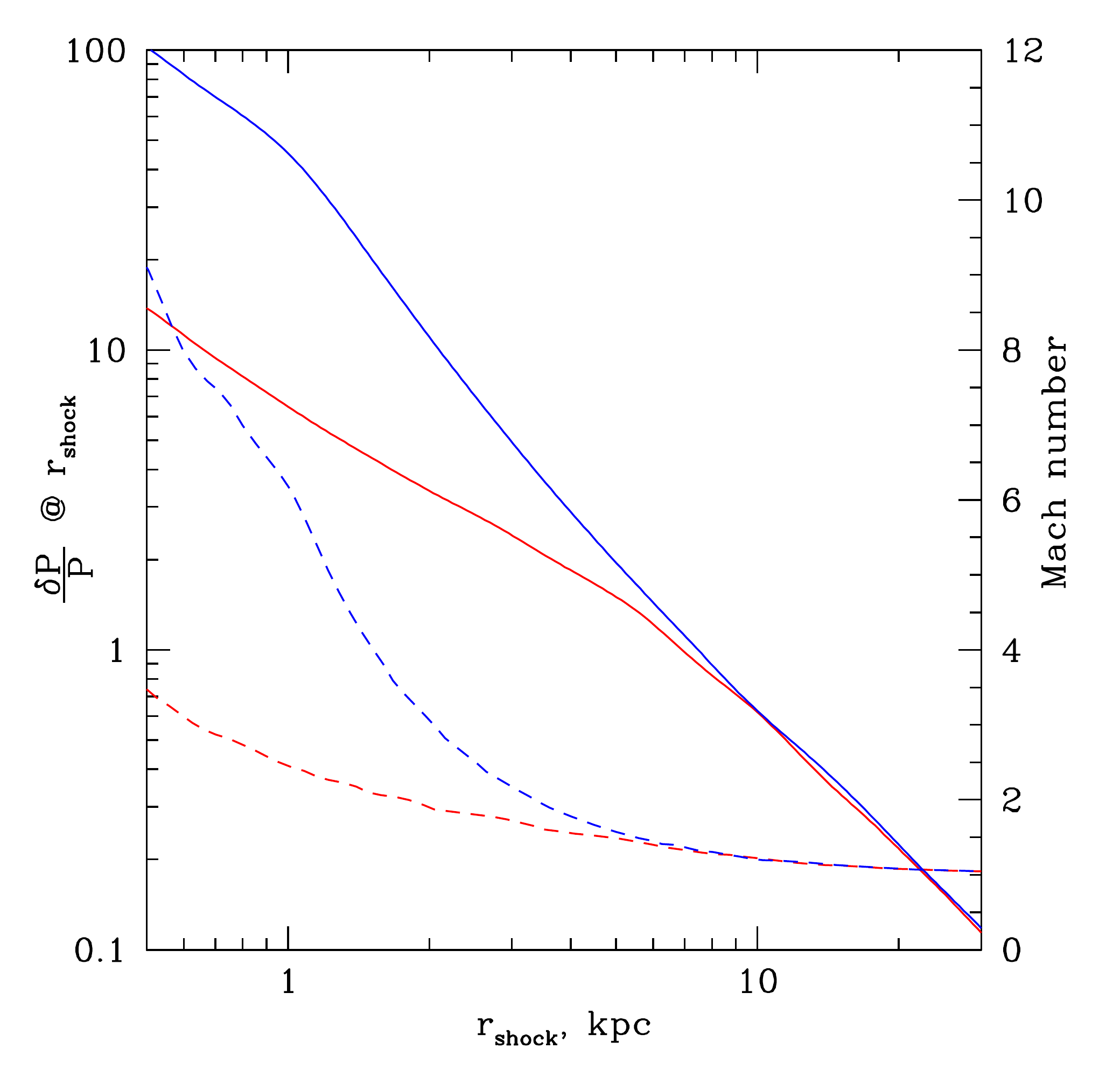}
}
  \caption{{\bf (a - left)} The gas density and gas temperature
    profiles resulting from two outbursts - one with a ``short''
      of 0.1 Myr duration and a second with a 2.2 Myr duration (the
      fiducial model) (blue and black curves respectively). The
    models are shown at the times that the modeled shock reaches 13
    kpc, the radius of the observed shock.  These times are 11 and 12
    Myr, for the shorter and longer/fiducial duration outbursts,
    respectively. The dashed lines (blue and black) indicate the
  outer radii of the piston that created the shock (marked with the
  red dashed line at 13 kpc). The piston from the short duration
  outburst is surrounded by a low density, hot shocked envelope.  The
  longer duration outburst (the fiducial model) is instead surrounded
  by a dense, cool envelope. At radii near the shock, the densities
  and temperatures of the two models are nearly identical. {\bf (b -
    right)} The evolution of the shock strength, parameterized as the
  pressure jump for the short and longer duration models. The
  pressure jump for outbursts of duration 0.6~Myr (blue, solid curve)
  and 2.2~Myr (red, solid curve) with outburst energy of $5.5\times
  10^{57}$~ergs is shown as a function of radius.  The pressure jumps
  for the two models differ dramatically at small radii where the
  Sedov-like outburst yields a much stronger over-pressure.  As the
  shocks evolve, they both match the observed shock at 13~kpc, but the
  Sedov-like outburst leaves a residue of hot,  strongly shocked
  gas as shown in panel (a).  For both outbursts, the equivalent Mach
  numbers are shown as dashed lines with axis on the right.  The upper
  dashed line (blue) is for the 0.6~Myr duration outburst and that for
  the 2.2 Myr outburst is the lower dashed (red) curve.}
\label{fig:slow-fast}
\label{fig:pjump}
\end{figure*}

\subsection{Short and Long Duration Outbursts}

To further illustrate the principles that drive the models described here and
how the duration of the outburst affects the appearance of the hot
corona, we select  two examples that illustrate the effects of the
  outburst duration on the properties of M87 -- a short duration
outburst and a longer duration outburst.  The short duration outburst
has a duration $\Delta t = 0.1\times10^6$ yrs while the longer
duration outburst has $\Delta t = 2.2\times10^6$ yrs (the blue and
black curves in Fig.~\ref{fig:duration}b).
Fig.~\ref{fig:models-schematic} shows graphically the dramatic
difference that may arise from the two different duration
  outbursts. 

Quantitatively, the different characters of the short and long
outbursts are shown in Fig.~\ref{fig:slow-fast}a where we label the
different regions that characterize the different types of
outbursts. We show the gas density and gas temperature profiles of the
$0.1\times10^6$~yr duration outburst (blue) and the fiducial
$2.2\times10^6$~yr duration outburst (black).  We have labeled the key
regions -- the piston, the hot, low density shocked envelopes (blue
text) for the short duration outburst and the piston and the cooler,
denser envelope for the fiducial duration outburst (black text).  
  Although the physics of the outbursts are identical, the duration
  imprints a qualitatively different signature on the surrounding
  atmosphere with quite different over-pressures and Mach numbers as a
  function of time (see Fig.~\ref{fig:pjump}b). For a given shock
  strength, the longer outburst produces a larger cavity, by a factor of
  three in volume, that can be used as a proxy for the outburst
  duration. 

The models are shown at the time when the modeled shock reaches
13~kpc.  For the two example outbursts (0.1 and 2.2~Myr durations)
considered in Fig.~\ref{fig:slow-fast}, the outburst ages (time for
the shock to reach 13 kpc) change by only  about 10\% (11 vs. 12 Myr
for the 0.1 and 2.2~Myr durations).  Despite the large difference in
initial Mach number (Fig.\ref{fig:pjump}b) for the outburst
energy ($5.5\times 10^{57}$) that yields density and temperature jumps
consistent with the observations, the age is dominated by the late
phases as the shock approaches 13~kpc.

Also, longer outbursts could be characterized by the absence of a hot,
low density envelope around the central cavity that is filled with
relativistic plasma. Such an envelope, characteristic of short
outburst models, is formed by the gas that has passed through the
strong shock. The lack of such an envelope in the data is consistent
with the ``long outburst scenario''. Whether it can be used as a
strong argument against the short outburst model depends on the
efficiency of thermal conduction in the gas, which is an open issue. 

\subsection{The Fiducial Model - a single outburst model for the 13 kpc
  shock}
\label{sec:gamma}

To quantitatively bound the family of outburst parameters, we examine
an ensemble of shock models where we have varied the outburst energy
and duration.  As described above, we first simulate the primary
outburst that produced the 13~kpc shock and assume, for this initial
comparison of observations to models, that this is the only outburst
that affects the inner 13~kpc of M87.  Our outburst model is  characterized by two key
outburst parameters - the duration (we assume constant power during the
outburst event) and the total energy deposited.

The parameters we must match are (a) the shock jump conditions which,
as noted above, primarily determine the total outburst energy
($E_{tot}$), and (b) the radius ($r_p$) of the radio cocoon, the
piston driving the shock, and (c) the observed radius of the 13 kpc
  shock. We could use either the temperature jump or the density jump
  to constrain the model.  The density jump is statistically more
  accurate but has a systematic uncertainty associated with the steep
  density gradient arising from the ``cool core'' atmosphere
  surrounding M87. The temperature jump is less accurate statistically
  but may provide a more realistic measure of the uncertainties
  inherent in the complex atmosphere of M87. Mach numbers derived for
  the density and temperature jumps are fully consistent (see Forman
  et al. 2007).  For the purpose of constraining the model parameters,
  we choose the less constraining temperature jump to better allow for
  the systematic uncertainties.

\begin{figure} 
  \centerline{\includegraphics[height=0.95\linewidth]{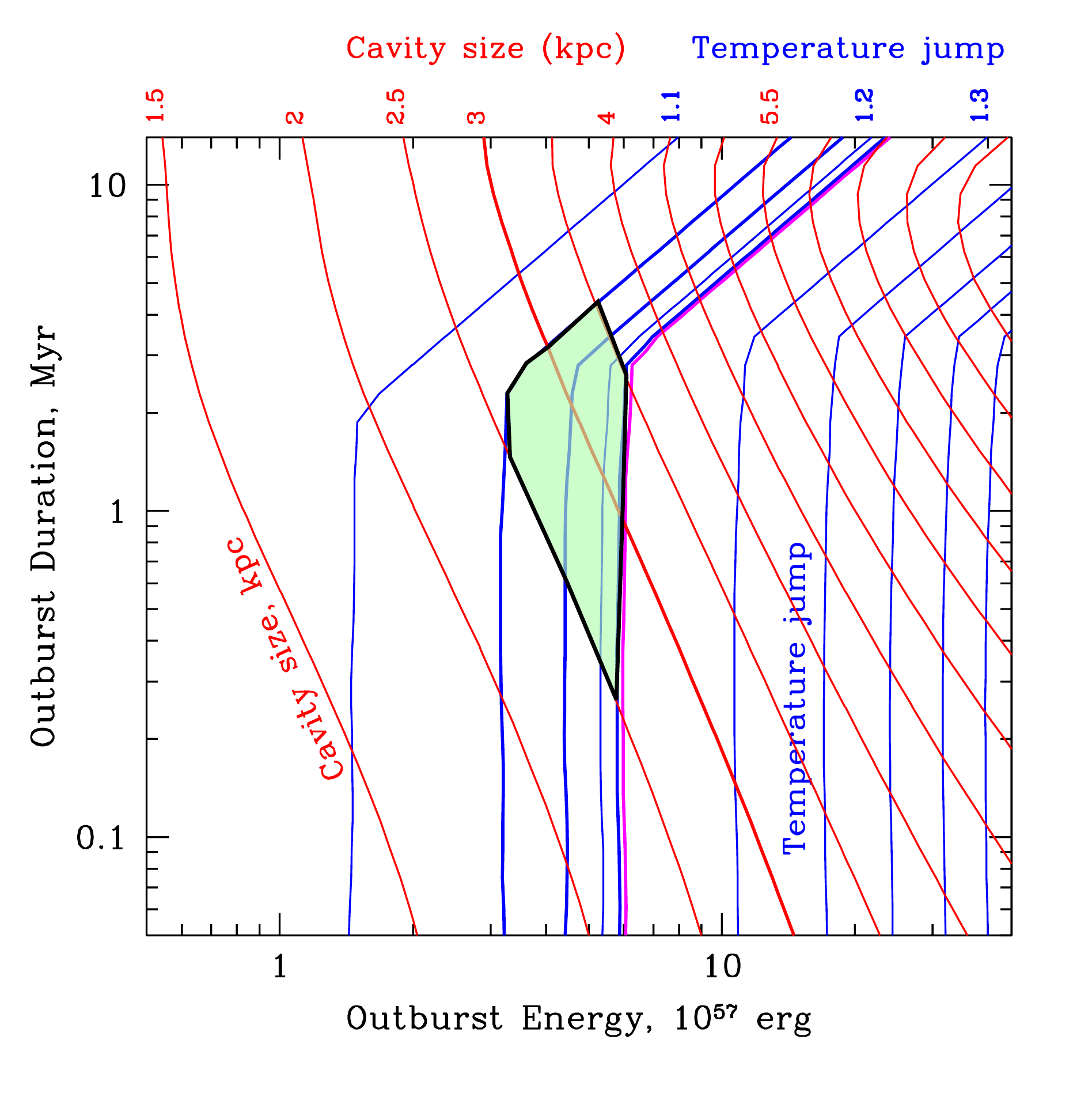}}
  \caption{ A grid of models as a function of outburst energy and
    outburst duration, for a one dimensional outburst model.  The
      model parameters are taken at the time when the modeled
      shock reaches 13 kpc, the radius of the observed shock.  Within
    this grid of models, we draw lines of constant temperature jump
    ($kT_{shock}/kT_{initial}$; blue solid lines) and constant
    piston size (radio cocoon; red solid lines). The values of the
    temperature jump and piston size are labeled along the top axis of the
    figure in the corresponding color.  The green region
    indicates the intersection between regions defined by
    $kT_{shock}/kT_{initial} = 1.18\pm0.03$  and cavity size of
    $3\pm0.5$~kpc. }
\label{fig:energy-duration}
\label{fig:grid}
\end{figure}

Fig.~\ref{fig:energy-duration} is a grid of models for two
  parameters -- the outburst energy, $E_{tot}$ and the outburst
  duration, $\Delta t$.  Loci of equal shock temperature jump (blue) and
  equal cavity size (red) are drawn.  The value of the gas temperature
  jump is $kT_{shock}/kT_{initial} = 1.18\pm0.03$ (Forman et
  al. 2007).   The second constraint arises from the size of the
  central cavity, the piston.   We identify the piston with the central
  radio cocoon which is labeled in the X-ray image shown in
  Fig.~\ref{fig:models-schematic}b (central panel) as well as in
  Fig.~\ref{fig:xray-radio}b and \ref{fig:shock-image}b.

As noted above and shown in Fig.~\ref{fig:energy-duration}, for
outburst durations less than about 3 Myr, the outburst energy is
{\em independent} of outburst duration (i.e., the loci of equal
density jumps are nearly vertical).  For durations longer than 3 Myr,
the acceptable range of energies does depend on the outburst duration.

The second constraint, the radius of the central cocoon, is
derived from the X-ray and radio images (see
    Fig.~\ref{fig:cavity-size-cylinder} as discussed in
    section~\ref{section:cavity-size}).  The intersection of the radius
    and density constraints indicates the most probable locus of points of
    (energy,duration) for the outburst.  The center of this region is
    $E_{tot}\sim5.5\times10^{57}$~ergs and $\Delta t\sim2$~Myr.

\begin{table} 
\caption{Fiducial Outburst Model in M87}
\begin{center}
\begin{tabular}{p{1.5in}c}  \hline   \hline
Outburst Age (Myr) &  12   \\
Outburst Duration (Myr)  & $\sim2$ \\
Outburst Energy ($10^{57}$ ergs)  & $\sim5$  \\ 
Energy carried by shock   & $\lesssim$22\%    \\ 
Thermal energy in cavity  & $\sim$27\% \\ 
Change in gravitational energy &  $\sim$40\% \\ 
Energy in shock heated gas    &  $\sim$11\% \\ \hline
Energy available for heating & $\sim$80\%   \\  \hline  \hline
\end{tabular}
\end{center}
\end{table}

With the known properties of the surrounding atmosphere and the
derived outburst details, we can compute the present epoch energy
partition arising from the outburst  (Table~1; see also Tang \&
  Churazov (2017) who ran a set of models with varying durations and
  energetics in a homogeneous medium to determine the energy partition
  and then mapped the results to more realistic density/temperature
  profiles.) For the fiducial outburst of $5.5\times10^{57}$ ergs,
  approximately 11\% of the energy resides in the kinetic energy of
  the shock (and a comparable amount in the thermal energy of the shock,
  since the shock is weak) that can be carried away from the central
region to larger radii since the shock is now relatively weak.  At least
50\% (and as much as 64\%) of the energy is contained in the enthalpy
of the central cavity/piston, and about 11\% of the energy has been
transformed into heating the gas as the shock moved outward to its
present position. In summary, in the
fiducial model, about 30\% of the outburst energy is deposited
in the shock.  In the model, about 10\% of this energy has already
been dissipated into heat as the shock traversed the region interior
to its present 13~kpc location.

Our 1D simulations
assume the adiabatic index $\gamma_g=5/3$ for the gas inside and
outside the ``piston''. If, in fact, the energy density inside the
piston is dominated by relativistic plasma with $\gamma_r=4/3$, the
thermal energy inside the cavity $\sim \frac{1}{(\gamma-1)}$ has to be
increased by the factor $\frac{\gamma_g-1}{\gamma_r-1}=2$ (see
Table~1), while keeping all characteristics of the gas outside the
piston unchanged. This would correspond to a moderate increase of the
total energy, required to inflate the bubble, and also a reduction in
the fraction of energy that goes into  the initial shock.

\begin{figure}
 \centerline{\includegraphics[height=0.95\linewidth]{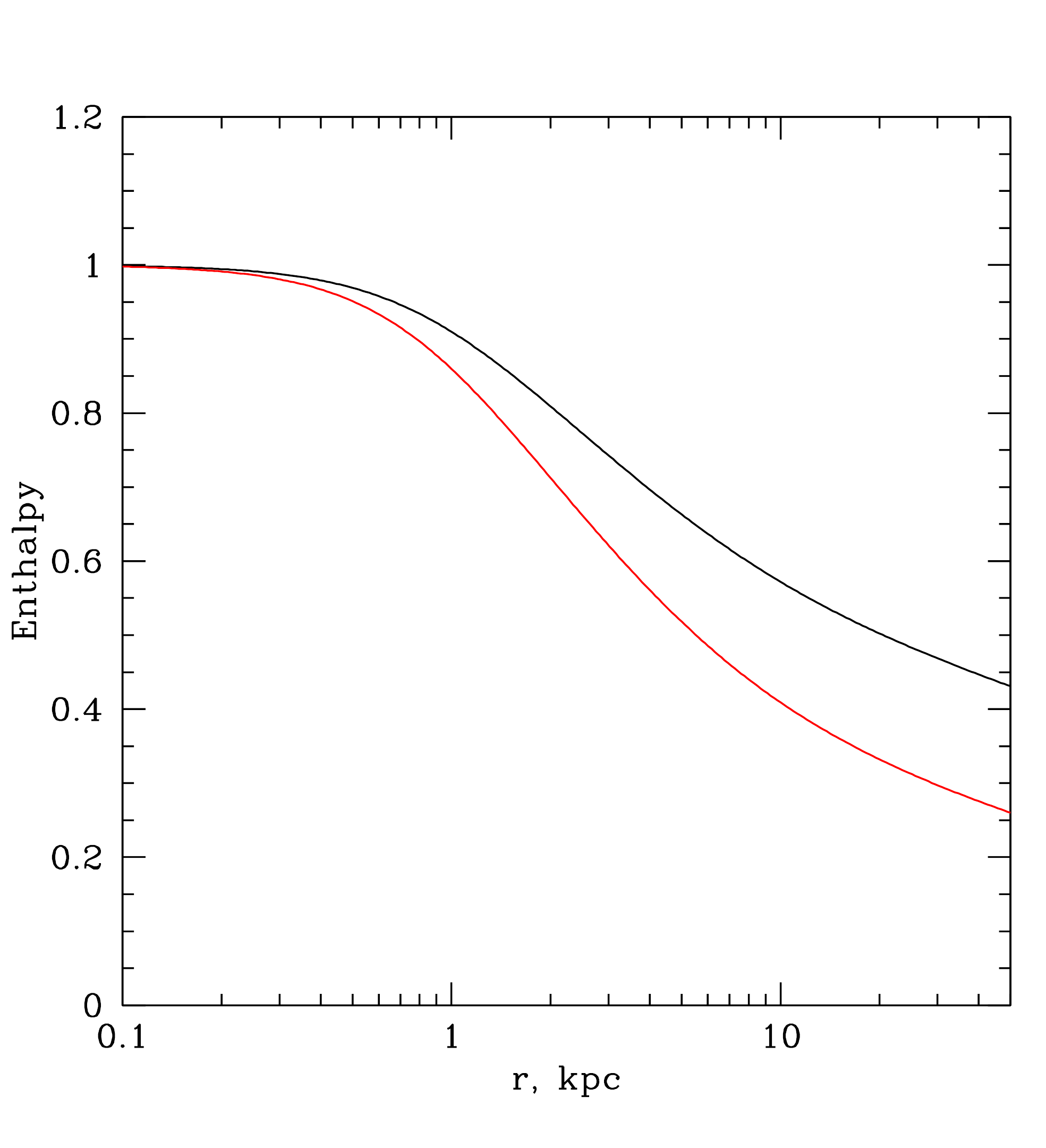}}
  \caption{For a buoyantly rising bubble, the fractional enthalpy loss for a plasma
  with adiabatic indices of 5/3 and 4/3 (upper and lower curves,
  respectively).  Buoyant plasma bubbles rising from the galaxy center
  to about 20 kpc, would lose approximately 50\% of their initial
  enthalpy which would ultimately be converted into
  thermal energy of the X-ray emitting plasma on a timescale that
  depends on the plasma microphysics.}
\label{fig:enthalpy}
\end{figure}

{\bf Enthalpy of the Central Bubble --}  The central bubble, the
  radio-emitting cocoon, contains a large fraction of the total
  outburst energy. Much of the enthalpy in a central bubble is
available for heating of the central region where radiative cooling is
important (e.g., see Churazov et al. 2001, 2002; see also Nulsen et
al. 2007). The fractional energy, $f$, retained by the buoyantly
rising bubble with adiabatic index $\gamma$, is given as
$f=(p_1/p_0)^{(\gamma-1)/\gamma}$ as the pressure changes from $p_0$
to $p_1$. For a relativistic plasma bubble, $\gamma=4/3$ and for a
non-relativistic plasma, $\gamma=5/3$. Fig.~\ref{fig:enthalpy} shows
the energy retained by a rising bubble in M87's atmosphere using the
fitted density and temperature profiles given in equations 1) and 2).
The enthalpy of the buoyant cocoon is dissipated into a variety of
forms including internal waves, sound waves, turbulent motion in the
wake of the bubble, potential energy of uplifted (cool) gas, and large
scale bulk flows.  While sound waves can carry energy away from the
central region, most other channels would eventually result in heating
the central region (see Churazov et al. 2001 for a more detailed
discussion on the containment of SMBH outburst energy in the core
region).  As Fig.~\ref{fig:enthalpy} shows, a buoyant bubble rising to
about 20~kpc in M87's atmosphere would lose about 50\% of its
enthalpy.  This energy will eventually be dissipated into heat on a
time scale that depends on the plasma microphysics.

\subsection{Multiple Outbursts}
\label{section:two}
\label{sec:two}

 The outburst that generated the 13 kpc shock is likely not the
  most recent one from M87's SMBH.  As the hard band images
  Fig.~\ref{fig:xray-radio}b and \ref{fig:shock-image}b show, there is
  a surface brightness enhancement surrounding the radio cocoon (the
  central bubble) indicating that the cocoon is an overpressurized
  region which is being driven by the current outburst we see in M87
  -- that also drives the existing jet.\footnote{ We note that there
    is an indication of a third weak intermediate age outburst with a
    surface brightness enhancement at $\sim1'$ (see
    Fig.~\ref{fig:surbri}) but we have not modeled
    this weak feature.} To understand the effects of the more recent
  outburst on our derived shock parameters, we add a second ongoing
  outburst at the present epoch to provide the observed overpressure
  within the central cocoon.

\begin{figure} 
  \centerline{\includegraphics[width=0.95\linewidth]{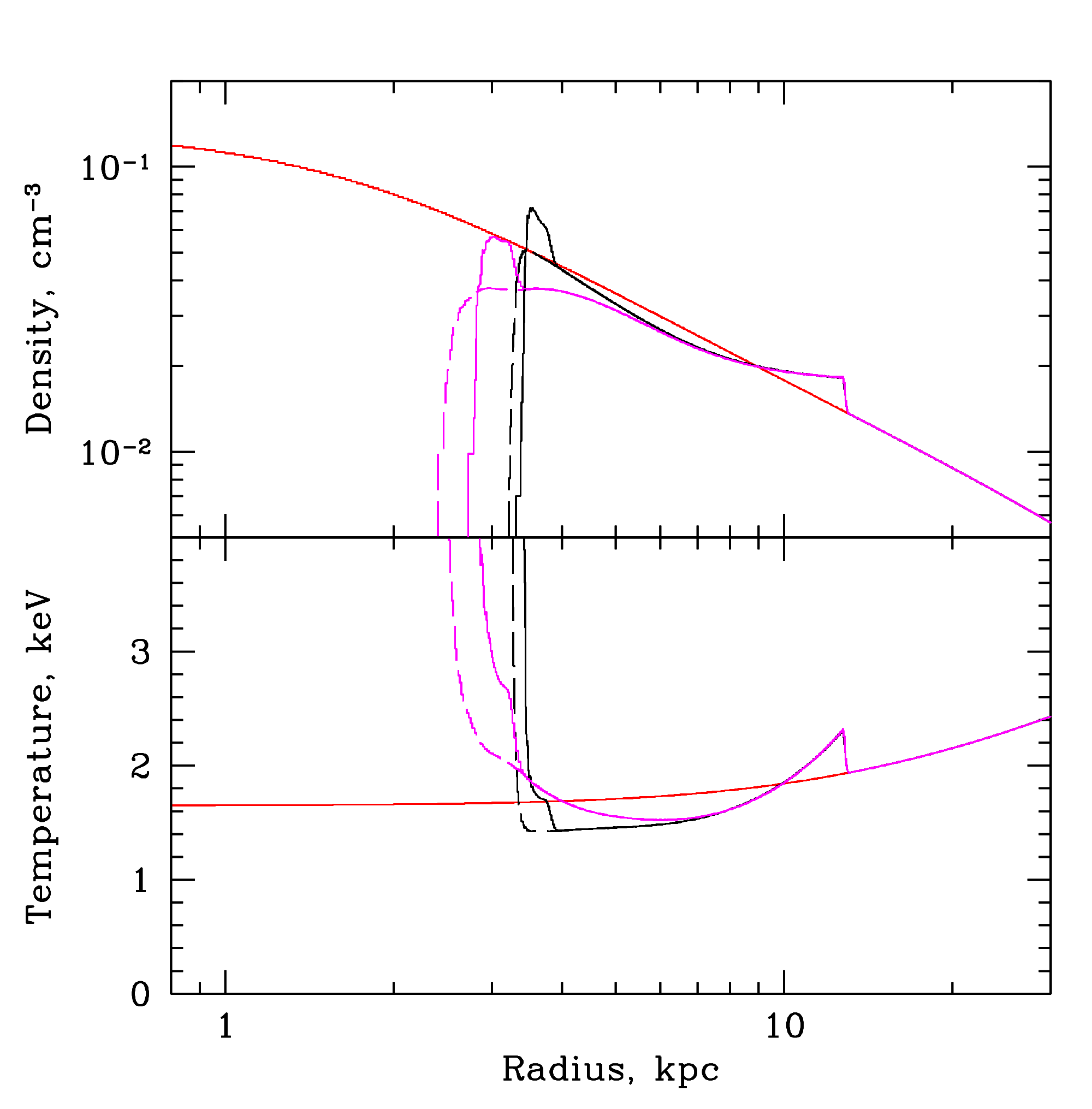}}
  \caption{Comparison of models with one and two outbursts.  We
    compare two single outburst models with energies of
    $E_{tot}=5.5\times10^{57}$~ergs but with long (2.2~Myr) and short
    (0.4~Myr) durations to models with a second outburst of cumulative
    energy $2\times10^{57}$~ergs and duration 1~Myr that is still
    ongoing.  The models, with differing ages, are shown at the times when the modeled
      shock reaches 13 kpc, the radius of the observed shock.  The
    single outburst models are shown with dashed lines.  The outburst
    models with a short primary outburst are in magenta and those with
    long outbursts are in black.  The addition of a second outburst
    injects energy into the existing cavity and makes only a small
    change to the predicted profiles.  To match the observations, the
    age of the main outburst must be reduced to 11~Myr, while its
    energy is unchanged.  }
\label{fig:multiple}
\end{figure}

The current
(ongoing) outburst has an energy (up to the present) of
$2\times10^{57}$~ergs (determined by the weak density jump at $\sim3$~kpc)
and a duration of about 1~Myrs.  If we include this recent outburst,
the age of the main outburst that produced the 13~kpc shock is reduced,
since the cavity size is slightly increased by the current
outburst. The presence of a second outburst reduces the {outburst age in} the
fiducial model by about 10\% to 11~Myrs.

With the above set of parameters, we find the gas density and gas
temperature profile shown in Fig.~\ref{fig:multiple}.  The figure
shows the central hot, low density cocoon which acts as the
piston. Just beyond the piston is the over-pressurized shell extending
to about 4 kpc. In our simple one dimensional model, the presence of an existing
cavity at the onset of the second outburst reduces the effects of an
initial short period of strong shock heating that might otherwise be
present at the beginning of the second outburst.  In 3D, if given
sufficient time between outbursts, the second outburst will encounter
a denser environment as the low density plasma rises buoyantly and is
displaced by denser plasma.  For short intervals between outbursts,
subsequent outbursts will have the effects of their initial expansion mitigated by
residual, low density plasma.

\subsection{The Central Piston}

The only large cavity that is seen interior to the 13~kpc shock is the
central cavity, the radio cocoon (Fig.~\ref{fig:xray-radio}b and
\ref{fig:shock-image}b).  Hence, this $\sim3$~kpc bubble  (equivalent 1D
size of the 3D bubble) of
relativistic plasma must be the piston that drove the 13~kpc shock.
However, since the relativistic plasma is buoyant, it will tend to
bifurcate into a dumbbell shape and each half will buoyantly rise.  Is
the presently observed cavity surrounding the M87 nucleus and the jet
consistent with having been created about $\sim11$~Myr ago when the
shock, presently seen at 13~kpc, was first created?  Churazov et
al. (2001) simulated the rise of buoyant bubbles in the M87/Virgo
system. They found a buoyant velocity over a wide range of radii of
about half the sound speed, $v_b\sim c_s/2$. In the M87 core, the gas
temperature is about 1~keV, giving a terminal buoyant velocity of
about 250~\kms.  Over a time $t_b\sim11-12$~Myr, the age of the 13~kpc
shock, the initial bubble will be pinched, form an elongated (possibly
dumbbell-like) shape and rise buoyantly to a distance $d_b \sim v_b
t_b$. The bubble system, at present, would therefore have dimensions
$\sim3 \times 8$~kpc, consistent with the  highly inclined jet angle
with respect to the line of sight (Biretta et al. 1999, Wang \& Zhou 2009) and
consistent with the 3D simulations presented in the next section (see
Figs.~\ref{fig:3d_map} and \ref{fig:3d_profiles}).

\subsection{3D Model of the AGN Outburst}
\label{sec:3d_simulation}
\label{sec:3d}

To test the sensitivity of our results to the simplifying
assumption of spherical symmetry, we performed 3D jet simulations to
replicate approximately the setup used in our 1D calculations.

Our simulations include a jet driven at a power of
$L_{\rm jet} = 1.2\times 10^{44}\,{\rm ergs\,s^{-1}}$ for
$\Delta t_{\rm jet} = 2\times 10^{6}\,{\rm yrs}$ into a $\beta$- model
atmosphere. Simulations were performed using the FLASH2.4 hydro code,
using the PPM solver (Fryxell et al. 2000), and following the same setup
described in Heinz et al. (2006) and Morsony et al. (2010). Simulations were
run with a central resolution of 50~pc and used AMR to focus
computational resources on the volume around the jet axis.

The simulations inject two oppositely directed jets, with the jet axis
random-walking within a cone of half-opening angle of ten degrees,
following the so-called dentist drill model (Scheuer 1982).

Consistent with the general model employed in this paper, the
expanding lobes excavate two cavities that drive an elliptical shock,
the semi-major axis of which is aligned with the mean jet
direction. The radio cocoon structure has a reasonable shape
compared to typical central cluster radio sources, with an aspect
ratio of approximately 3:1.

\begin{figure} [h]
\includegraphics[width=0.95\linewidth]{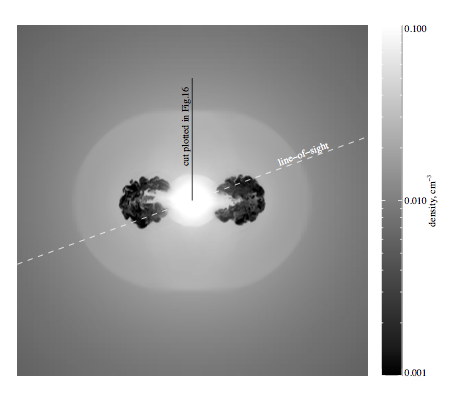}
\caption{Density slice through 3D simulation of the jet-driven shock
  in M87 (see \S\ref{sec:3d_simulation} for details of the
  setup). }
\label{fig:3d_map}
\end{figure}

Simulations were run until the semi-minor axis of the shock reached
the measured shock size in M87 of 13 kpc. A density slice through the
jet axis is shown in Fig.~\ref{fig:3d_map}, showing the
under-dense radio lobes and the shock. The aspect ratio of the shock
is approximately 1.3:1.

The jet viewing angle in M87 is likely close to the line of sight
(e.g., Biretta et al. 1999, Wang \& Zhou 2009). Thus the elongation of the
shock is likely hidden by projection. It is therefore appropriate to
use measurements of the shock properties along the semi-minor axis in
the simulation for comparisons with observations and the 1D
models. Because an elongated shock requires a larger energy (roughly
by the aspect ratio of 1.3) compared to a spherical shock, we used a
larger total injected energy of $E_{\rm jet} = L_{\rm jet}\Delta
t_{\rm jet} = 7.4\times 10^{57}\,{\rm ergs} = 1.3\times E_{1D}$.

\begin{figure}
\includegraphics[width=0.95\linewidth]{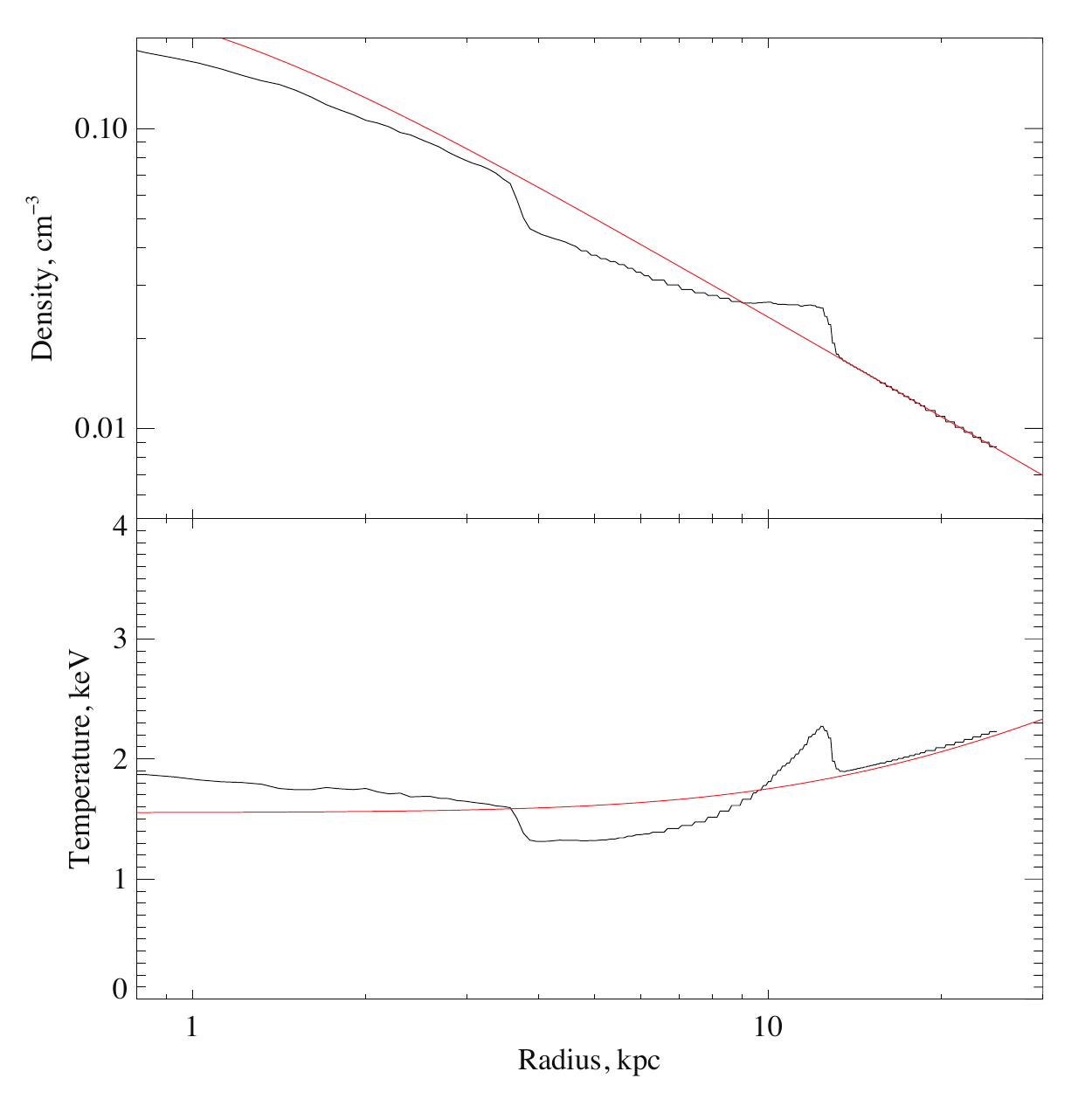}
\caption{Density and temperature profiles along the semi-minor axis of
  the shock in the 3D simulations  shown in Fig.~\ref{fig:3d_map}.  The
    ``edge'' feature seen at $\sim4$~kpc is produced by emission from
    gas that has refilled the volume (visible as light region in
    Fig.~\ref{fig:3d_map}) behind the expanding piston (visible as a
    dark region in Fig.~\ref{fig:3d_map}).  The depression in density
    that extends to $\sim8$~kpc arises from the lower density plasma
    in the expanding cocoon along the line of sight.  The decrease in
    temperature behind the shock is the characteristic of weak shocks
    discussed in Fig.~\ref{fig:fiducial-model}.}
\label{fig:3d_profiles}
\end{figure}

The radial density and temperature profiles along the semi-minor axis
of the shock are plotted in Fig.~\ref{fig:3d_profiles}.  Outward of
the 1D piston location, they agree well with the profiles plotted in
Fig.~5 for our fiducial 1D model. In particular, the density and
temperature jumps at the shock agree well with the 1D model,
supporting the use of these measurements as observational
diagnostics. The simulations also show a low-temperature post-shock
region between 3.5 kpc and the shock, as predicted by the 1D fiducial
model. Furthermore, the 3D model reproduces the distinguishing
characteristic of the piston-driven expansion:  the absence of a
  large increase in temperature outside the piston (and interior to
  the shock) that would be produced by impulsive (instantaneous)
energy injection in a Sedov-like mode. We note again that this relies
on the assumption that thermal conduction is negligible.

The main difference between our 1D and 3D models is the presence of
{\em two} off-center pistons in the 3D case, which leads to the
elongation of the shock. The central region in the simulation is
re-filled by gas that falls back and generates a new (slightly hotter)
core after the jet is turned off.

It is straightforward to understand why the 1D model is so successful
in reproducing the properties of the 3D simulations in the direction
perpendicular to the jet: The lateral expansion of a cocoon-driven
shock is energy (i.e., pressure) driven, while the initial
longitudinal expansion of the shock is driven by the momentum of the
jet, as argued by Begelman \& Cioffi (1989). The only correction required
between 1D and 3D is therefore the total volume of the shock, which
increases the energy  (by about 30\%) required to drive a shock to a given semi-minor
axis by the shock aspect ratio, as confirmed by our simulations.

\section{Conclusion}

Hot gaseous atmospheres are ideal tracers of major events in the
evolution of brightest cluster (or group) galaxies (BCGs), their
central supermassive black holes (SMBHs), and their dark matter halos.
In addition to evidence of outbursts, X-ray images and temperature
maps provide constraints on gas mixing from mergers through shocks,
cold fronts, and ``gas sloshing'' (e.g., Markevitch \& Vikhlinin 2007;
Markevitch, Vikhlinin \& Mazzotta 2001, Markevitch et al. 2002;
Vikhlinin, Markevitch \& Murray 2001; Johnson et al. 2010).  Abundance
distributions also show evidence for gas motions and merging events
(e.g., Rebusco et al. 2005, 2006; Xiang et al. 2009; Simionescu et
al. 2010). Another ICM tracer of the dynamic history is encoded in the
X-ray surface brightness fluctuations (Churazov et al. 2012,
Zhuravleva et al. 2014).  For M87, the obvious outburst history
extends over about 100~Myrs. Our discussion above has concentrated on
the outburst that produced the nearly circular shock at 13~kpc and the
central ``bubble'' whose inflation drove the shock into the
surrounding atmosphere.  The relatively simple geometry of the system
provided the opportunity to explore the details of the outburst and
yielded quantitative estimates of the outburst properties including
its age, $\tau\sim11-12$~Myrs, its energy,
$E_{tot}\sim5-6\times10^{57}$~ergs, and duration,
$\Delta_t\sim1-3$~Myrs.  In addition, we are able to estimate the
present epoch energy partition with about 80\% of the energy available
for heating the gas and about 20\% carried away,  beyond 13~kpc,
  by the shock as it weakens to a sound wave (see Table~1). Thus,
  during the outburst, in the fiducial model, about 30\% of the
  outburst energy is deposited in the shock.  In this model, $\sim$
  10\% of this energy has already been dissipated into heat as the
  shock traversed the region interior to its present 13~kpc location,
  while the remaining $\sim 20$\% is carried to larger radii.  For
M87, a large fraction of the outburst energy resides in the central
bubble enthalpy.  As Churazov et al. (2001, 2002) argued, the bulk of
this energy is converted to thermal energy of the X-ray emitting gas
in the central region surrounding M87.

In the context of our simple model, we also are able to estimate the
properties of the current, ongoing outburst that has only slightly altered the
signature left by the preceding outburst.  The signature of the
current outburst is consistent with having begun about 1~Myr ago and having
injected $2\times10^{57}$~ergs into the preexisting cavity.  As noted
above, the $\sim11-12$~Myr old outburst inflated a cavity that is now
elongated, at least partially by buoyancy.  While the exact values
describing the M87 outbursts are uncertain, with the outburst energy
somewhat larger than the 1D model predicts (see
section~\ref{sec:3d_simulation} and the discussion of the 3D model),
the qualitative description of a ``slow''  (few Myr) outburst
remains valid and is confirmed by the more realistic 3D model.

 M87 provides a view of a
``typical'' outburst from a low-Eddington rate accretor with the bulk
of the energy liberated as mechanical, rather than radiative, energy.
Considerable attention has been given to the very energetic outbursts
in luminous clusters (e.g., Nulsen et al. 2005) and to some of the
spectacular outbursts in groups (e.g., NGC5813; Randall et
al. 2011). However, luminous early type galaxies also have hot coronae
(Forman, Jones \& Tucker 1985) and, like their more luminous cousins,
also harbor mini-``cooling cores''.  In the absence of any heating,
these systems would have mass deposition rates up to a few solar
masses per year (Thomas et al. 1986) and yet they host very little
star formation and remain ``red and dead'' (e.g., Hogg et al. 2002).
Outbursts very similar to those discussed for M87 are also present in
these systems.  NGC4636 (Jones et al. 2002, Baldi et al. 2009), M84
(Finoguenov et al. 2008), and NGC4552 (Machacek et al. 2006) are
representative examples of this class.

There are a variety of energy sources suitable for replenishing the
radiated energy from the hot gas in galaxy cluster cores. Two of the
most prominent are mergers and AGN outbursts which drive gas
motions. In M87, we see effects of both processes, e.g., a) ongoing
mergers such as M86 (Forman et al. 1979, Randall et al. 2008) and gas
sloshing of the entire Virgo core (Simionescu et al. 2010) and b) AGN
outbursts from M87 as we have discussed in detail above that inflate
buoyant bubbles.  In the context of gas sloshing, ZuHone et
  al. (2010) have discussed the mixing of hotter gas from larger radii
  with cooler gas from the central regions of the cluster (or
  galaxy). The mechanism for converting the bulk motions to heat has
only recently been probed. Zhuravleva et al. (2014) argued that
gas motions, that arise from both merging and SMBH feedback
(primarily, motions driven by the rise of buoyant plasma bubbles as
discussed for M87 above), are very likely converted to thermal energy
via dissipation of turbulence.  The turbulent heating inferred for M87
(and Perseus) is sufficient to balance the radiative cooling.  Hence,
we can now begin to quantitatively understand the feedback process and
conversion of gas motions to thermal energy of the gaseous atmosphere.

 The outbursts from M87 are characteristic of radiatively inefficient
accretion (e.g., Ichimaru 1977, Rees et al. 1982, Narayan \& Yi 1994,
Abramowicz et al. 1995, Blandford \& Begelman 1999, Yuan \& Narayan
2014). Early-type galaxy evolution models that include both radiative
and mechanical feedback have been explored extensively.  Pellegrini,
Ciotti \& Ostriker (2012, and references therein) have modeled the
evolution of isolated early type galaxies over cosmological
times. They find episodic outbursts with high quasar-like radiative
luminosities ($\sim10^{46}$~\ergss) at early epochs.  M87, and most
present epoch early-type galaxies, lie in richer environments (cluster
or group centers or cluster cores).  Although the gas environment is
much richer, present epoch early-type galaxies appear to have more
moderate outbursts than those at earlier epochs. Future X-ray missions
will be able to study the detailed properties of outbursts and probe
the conversion of bulk motions to thermal energy (e.g., Croston et
al. 2013, Vikhlinin 2012).  The ability to probe to high redshift with
arc second angular resolution (Vikhlinin 2012 see section 3.1 and
Fig.~3) could fully test models of galaxy evolution and the impact of
the SMBHs that lie in their nuclei, by tracing the evolution of both
the AGN and the surrounding hot gaseous atmosphere and deriving
properties (luminosity, temperature, density profile, and abundance)
from redshifts of $z\sim6$ to the present.

\acknowledgments

This work was supported by contracts NAS8-38248, NAS8-01130,
NAS8-03060, the Chandra Science Center, the Smithsonian Institution,
the Institute for Space Research (Moscow)
and Max Planck Institute f\"{u}r Astrophysik (Munich). S.H. acknowledges
support through NSF grant AST 1109347.  We thank the anonymous referee
whose comments significantly improved the paper.



\end{document}